\documentclass[%
 reprint,
 superscriptaddress,
 amsmath,amssymb,
aps,
pra,
]{revtex4-2}

\usepackage{CJKutf8}
\usepackage{lineno}
\usepackage[colorlinks=true,linkcolor=blue]{hyperref}

\usepackage[squaren]{SIunits}
\usepackage{color}
\usepackage{amsmath}
\usepackage{caption,subcaption}
\usepackage{setspace}
\usepackage{multirow}
\usepackage{threeparttable}
\usepackage{booktabs}
\usepackage{graphicx}
\usepackage{dcolumn}
\usepackage{bm}
\usepackage{rotating}
\definecolor{americanrose}{rgb}{1.0, 0.01, 0.24}
\definecolor{coralpink}{rgb}{0.97, 0.51, 0.47}
\definecolor{ao(english)}{rgb}{0.0, 0.5, 0.0}
\definecolor{darkpastelgreen}{rgb}{0.01, 0.6, 0.2}
\definecolor{cyan(process)}{rgb}{0.0, 0.72, 0.92}
\definecolor{brown}{rgb}{0.8, 0.5, 0}

\begin{document}
\preprint{APS/123-QED}

\title{
 Franck-Condon Simulation of Vibrationally-Resolved X-ray Spectra for Diatomic Systems: Validation of Harmonic Approximation and Density Functional Theory
}

\author{Lu Zhang}
 \affiliation{MIIT Key Laboratory of Semiconductor Microstructure and Quantum Sensing, Department of Applied Physics, School of Physics, Nanjing University of Science and Technology, 210094 Nanjing, China}
 
 \author{Minrui Wei}
 \affiliation{MIIT Key Laboratory of Semiconductor Microstructure and Quantum Sensing, Department of Applied Physics, School of Physics, Nanjing University of Science and Technology, 210094 Nanjing, China}
 
 \author{Guoyan Ge}
 \affiliation{MIIT Key Laboratory of Semiconductor Microstructure and Quantum Sensing, Department of Applied Physics, School of Physics, Nanjing University of Science and Technology, 210094 Nanjing, China}
  
 \author{Weijie Hua}%
 \email{wjhua@njust.edu.cn}
 \affiliation{MIIT Key Laboratory of Semiconductor Microstructure and Quantum Sensing, Department of Applied Physics, School of Physics, Nanjing University of Science and Technology, 210094 Nanjing, China}
 
\date{\today}


\begin{abstract} 
Under the Franck-Condon approximation, we systematically validated the performance of density functional theory (DFT) and the effects of anharmonicity in simulating C/N/O K-edge vibrationally-resolved X-ray spectra of common diatomic molecules. To get ``transparent'' validations, vibronic fine structures of only the lowest 1s excited or ionized state in the  X-ray absorption (XAS) or photoelectron (XPS) spectra were investigated. All 6 systems (N$_2$, N$_2^+$; NO, NO$^+$; CO, CO$^+$)  were studied within the harmonic oscillator (HO) approximation using DFT with four functionals (BLYP, BP86, B3LYP, M06-2X) for 10 XAS and 4 XPS spectra, and excellent agreement between theoretical and experimental spectra was found in most systems, except O1s XAS of NO, CO, and NO$^+$.  We analyzed and established a connection between their complex vibronic structures (many weak oscillating features within a broad peak) and the significant geometrical changes induced by the O1s hole.  The three spectra were well reproduced with anharmonic (AH) calculations by using quantum wavepacket dynamics based on potential energy curves (PECs) generated by DFT methods or multiconfigurational levels, highlighting sensitivity to the anharmonic effect and the PEC quality. In other examples of XAS (CO$^+$, C1s and O1s; NO, N1s) corresponding to smaller structural changes,  HO and AH  approaches lead to similar fine structures, which are dominated by 0-0 and 0-1 transitions. This study highlights the use of DFT with selected functionals for such diatomic calculations due to its easy execution and generally reliable accuracy. Functional dependence in diatomic systems is generally more pronounced than in polyatomic ones. We found that BLYP, BP86, and B3LYP  functionals consistently exhibited high accuracy in predicting spectral profiles, bond lengths, and vibrational frequencies, which slightly outperformed M06-2X.
\end{abstract}

\maketitle

\clearpage
\section{Introduction}
Spectroscopy serves as a crucial link between experimental observations and theoretical understanding in physics, with precision being the key factor. One notable example in the history of physics is the hydrogen spectrum, where increasing measurement precision has reciprocally fostered the development of quantum theory. High-resolution spectroscopy acts as a litmus test for assessing the precision of the theoretical methods employed. Particularly, in the context of high-resolution vibrationally-resolved X-ray spectroscopy,\cite{carravetta_x-ray_2022, hergenhahn_vibrational_2004, svensson_soft_2005, book_ESCA_molecules, gelmukhanov_theory_1977, chen_k_1989, rennie_comprehensive_2000, minkov_naphthalene_2004, hoshino_vibrationally_2008, fronzoni_vibrationally_2014, mosnier_inner-shell_2016, vaz_da_cruz_anomalous_2018,  michelitsch_efficient_2019, Thesis_Kjellsson_2021, bilalbegovic_sulfur_2021, huang_theoretical_2022, schippers_vibrationally_2023} comparison to experimental data affords an evaluation of the quality of potential energy surfaces (PESs) for both the core-electron excited or ionized (in the subsequent context, we will use the term ``\textit{excited state}'' to refer to either case without distinction) and the ground electronic states. It also provides insights into the accuracy of the electronic structure methods used, especially for states with a core hole. In addition to their conceptual significance, high-resolution X-ray spectra of molecules and ions have important implications in various fields, including X-ray astronomy and interstellar chemistry.\cite{bilalbegovic_sulfur_2021, zeegers_dust_2019, frati_oxygen_2020} 

Accurately simulating detailed X-ray spectra helps analyze and uncover physical and chemical insights. Success relies on considering both vibronic coupling (VC) and the electronic structure method used. Various VC models are distinguished by the approximations and theoretical frameworks (i.e., time or frequency domains) they employ. The harmonic oscillator (HO) approximation is the most fundamental approximation widely used in simulations of vibrationally-resolved electronic spectroscopy in different regimes as well as vibrational spectroscopy. It is particularly advantageous for polyatomic molecules and is usually a reliable approximation. When studying vibronic fine structures in X-ray spectra, the harmonic approximation is commonly adopted alongside the time-independent (TI) framework,\cite{cerezo_fcclasses3_2023} allowing for a clear analysis of individual transitions.  Successful computations\cite{hua_theoretical_2020, wei_vibronic_2022, wei_vibronic_2023, wei_vibronic_2023_iii, cheng_vibrationally-resolved_2022} for polyatomic molecules and ions have been achieved through Franck-Condon (FC) simulations that consider the Duschinsky rotation (DR) effect.\cite{duschinsky_interpretation_1937} On the other hand, a more advanced approach is to consider the anharmonic (AH) effect based on accurately generated PESs. This is usually used for relatively small systems, for instance, diatomic\cite{couto_anomalously_2016, zhang_nonlinear_2016} and triatomic\cite{couto_selective_2017} systems. Practical implementations are often associated in the time-dependent (TD) wavepacket framework, by directly solving the TD Schr\"{o}dinger equation (TDSE),\cite{couto_anomalously_2016, couto_selective_2017} or by using efficient and approximate dynamical simulations, for instance, the multiconfigurational time-dependent Hartree (MCTDH)\cite{beck_multiconfiguration_2000} calculations.\cite{zhang_nonlinear_2016}    


In terms of the electronic structure method, density functional theory (DFT) has emerged as a predominant method that offers profound insights into molecules, surfaces, and materials, which well balances accuracy and efficiency. DFT with restrictions on the core orbital occupations has demonstrated the capability to accurately predict X-ray spectra.\cite{huareview_2012, zhang_nonlinear_2016, norman_simulating_2018, besley_density_2020, hua_structure_2011}  It has shown good agreement with experiments for vibrationally-resolved X-ray photoelectron (XPS)\cite{hua_theoretical_2020, wei_vibronic_2022, wei_vibronic_2023, cheng_vibrationally-resolved_2022} and absorption (XAS)\cite{couto_breaking_2021} spectra of a comprehensive spectrum of molecules, such as benzene,\cite{hua_theoretical_2020} furan,\cite{hua_theoretical_2020} pyridine,\cite{hua_theoretical_2020} azines,\cite{wei_vibronic_2022} indoles,\cite{wei_vibronic_2023} polycyclic aromatic hydrocarbons (PAHs),\cite{cheng_vibrationally-resolved_2022} and ions like N$_2$H$^+$,\cite{couto_breaking_2021}  when it was combined with FC simulations including the Duschinsky rotation\cite{duschinsky_interpretation_1937}  effect. In these calculations, either full  (FCH)\cite{triguero_calculations_1998} or excited (XCH)\cite{prendergast_x-ray_2006}  core-hole approximation was leveraged to describe the core ionized (in the case of XPS) or excited (in the case of XAS) states, while conventional DFT was employed to model the ground state (GS).  It is worth noting of other DFT applications\cite{minkov_naphthalene_2004, minkov_biphenyl_2005, hua_theoretical_2020} also with the half core-hole [HCH, i.e., transition potential (TP)]\cite{triguero_calculations_1998} or the equivalent core-hole (ECH, i.e., Z+1)\cite{jolly_thermodynamic_1970} approximations. Buoyed by the promising outcomes from DFT computations, we are currently engaged in formulating a theoretical library for vibrationally-resolved XPS/XAS spectra of common molecules and ions, with initial findings already disseminated.\cite{hua_theoretical_2020, wei_vibronic_2022,  wei_vibronic_2023, wei_vibronic_2023_iii, cheng_vibrationally-resolved_2022} Most investigated systems were polyatomic, i.e., composed of three or more atoms. 

Diatomic systems, with a single vibrational mode, offer a more ``transparent'' view of vibronic coupling effects. They are highly sensitive to the utilized vibronic coupling and electronic structure methods. Extensive investigations have been conducted on diatomic molecules such as N$_2$,\cite{ehara_symmetry-dependent_2006} CO,\cite{puttner_vibrationally_1999} NO,\cite{puttner_vibrationally_1999} and ions like NH$^+$,\cite{carniato_vibrationally_2020} N$_2$$^+$,\cite{lindblad_x-ray_2020} CO$^+$,\cite{couto_carbon_2020}  NO$^+$,\cite{lindblad_experimental_2022} and C$_2^-$\cite{schippers_vibrationally_2023} through both XPS and XAS experiments,\cite{ehara_symmetry-dependent_2006, puttner_vibrationally_1999, hergenhahn_vibrational_2004, carniato_vibrationally_2020, lindblad_x-ray_2020, couto_carbon_2020, lindblad_experimental_2022, schippers_vibrationally_2023} as well as theoretical simulations at various levels.\cite{carravetta_x-ray_2022, ehara_symmetry-dependent_2006, puttner_vibrationally_1999, carniato_vibrationally_2020, lindblad_x-ray_2020, couto_carbon_2020, lindblad_experimental_2022, schippers_vibrationally_2023} Generally, good agreement between experimental and theoretical results has been observed. Alongside technological advancements in X-ray synchrotron beams for experimental development\cite{mosnier_inner-shell_2016}, theoretical methods for X-ray spectra have also seen significant improvements in recent years,\cite{carravetta_x-ray_2022} enabling high-quality theoretical predictions of vibrationally-resolved X-ray spectra. Neutral diatomic molecules have been a frequent focus in many X-ray spectral studies due to their ability to reach higher target densities compared to charged particles.\cite{schippers_vibrationally_2023} Recent innovations in ion beam and ion trap techniques have facilitated precise inner-shell investigations involving positively charged diatomic molecular ions.\cite{lindblad_experimental_2022, carniato_vibrationally_2020}

Previously, various post-Hartree-Fock methods were used in almost all diatomic calculations.  Small systems allow advanced electronic structure methods. Couto et al.\cite{couto_carbon_2020} presented and analyzed high-resolution XAS spectra of CO$^+$  by employing wavepacket dynamics calculations based on potential energy curves (PECs) computed with the restricted active space second-order perturbation (RASPT2) method.  Lindblad et al.\cite{lindblad_experimental_2022} interpreted the experimental XAS spectra of N$_2^+$\cite{lindblad_x-ray_2020} and NO$^+$ using the restricted active space self-consistent field (RASSCF) and RASPT2 methods. Carniato et al.\cite{carniato_vibrationally_2020} simulated the PECs of low-lying N1s excited states of NH$^+$ by using the configuration interaction singles and doubles (CISD) method and the valence-state PECs by multiconfigurational self-consistent field (MCSCF). Rocha\cite{rocha_potential_2011} simulated the PEC of CO in the lowest C1s excited state by the RASSCF method. Martins et al.\cite{martins_disentangling_2021} computed PECs of low-lying F1s core excited states of HF and HF$^+$ by using the full configuration-interaction (FCI) method to help understand the photo-dissociation dynamics. Such studies have been instrumental in enriching our understanding of core-hole state electronic structure, X-ray physics, vibronic coupling, and bond dissociation dynamics. However, such studies typically focus on one or only a few systems, and computational studies by different groups often employ different theoretical levels. The lack of simulations for multiple systems on the same footing hinders fair comparisons and the derivation of general rules regarding VC properties.

The goal of the present study is (1) to utilize both the HO and AH methods in order to validate the influence of anharmonicity in simulating vibrationally-resolved X-ray spectra of diatomic systems, as well as (2) to validate the performance of the DFT method. The Condon approximation\cite{condon_theory_1926, *franck_elementary_1926} is always assumed in both approaches. First, with the HO approach, we aim to benchmark DFT in vibrationally-resolved XPS/XAS spectral simulations of common diatomic systems.  Despite potential limitations at larger bond distances where multiconfigurational effects are crucial, DFT has demonstrated the ability to generate accurate PECs near the FC region. The FCH/XCH approximation will be utilized for the lowest 1s ionized/excited state. XAS (N$_2$, N$_2^+$; NO, NO$^+$; CO, CO$^+$) and XPS (NO and CO) spectra of common molecules and ions, covering the C/N/O K-edges, will be considered. Second, this study will also present anharmonic studies for XAS spectra of selected systems (CO$^+$, NO, CO, and NO$^+$).  In addition to the four DFT methods, we will also compute the spectra with multiconfigurational methods for comparisons, for CO$^+$ and CO only,  with the input PECs generated by us or obtained from previous publications. It is noteworthy that most diatomic systems investigated in this study (except N$_2^+$), are molecules or ions detected in astronomical observations of interstellar or circumstellar environments.\cite{tielens_molecular_2013, wiki_interstellar} This underscores the importance of our research in the fields of X-ray astronomy and interstellar chemistry, which currently lack high-resolution spectra, particularly in the X-ray range, for reference purposes.


This work is also motivated by a previous study\cite{couto_breaking_2021} (by one of the coauthors, WH) on a triatomic system, N$_2$H$^+$, where DFT well predicted the double-well potentials of the lowest N1s excited states of both nitrogens (along one of the bending vibrational modes) and generated good XAS spectral agreement with the experiment. Thus, it is reasonable to anticipate good accuracy using DFT for diatomic systems, considering their even simpler PECs,  typically well-described by Morse potentials. Through the current comprehensive investigation of different VC models and electronic structure methods, valuable insights are expected to be provided for accurate and efficient simulations of high-resolution X-ray spectroscopy. 

\section{Methodology} \label{sec:method} 

\subsection{Notations}  
 
We use bolded symbols in lower and upper cases for vectors and matrices, respectively, while symbols in normal fonts indicate scalars. $g$ refers to the  initial (ground) electronic state, and $e$ stands for the final (excited) electronic state. To distinguish between quantities in the initial and final states, we use symbols with or without a prime. As such, the vibrational quantum numbers in both the initial and final states are denoted as $n'$ and $n$, respectively. The vibronic transition involved in the XPS/XAS process is expressed as:
\begin{equation}
|g,n'=0\rangle\rightarrow |e, n\rangle.
\end{equation}
Besides, the equilibrium geometries of the initial and final states are denoted by $\mathbf{x}'$ and $\mathbf{x}$, using Cartesian coordinates in column vectors. The normal coordinates for both states are represented by  $\mathbf{q}'$ and  $\mathbf{q}$, also in column vectors. The respective normal mode matrices are indicated by $\mathbf{L}'$ and  $\mathbf{L}$. Vibrational frequencies are designated by $\omega'$ and  $\omega$, while bond lengths are represented by $R'$ and $R$. 

The vibrational wavefunctions in ground state $g$ and excited state $e$ are denoted as $|\varphi^g_{n'}\rangle$ and $|\phi^e_n\rangle$, respectively. The corresponding vibrational energies are $\varepsilon^g_{n'}$ and $\epsilon^e_{n}$. Throughout the formulas that follow, atomic units (a.u.) are used, where $\hbar = 1$.

\subsection{Harmonic approach}  

The computational approach using the harmonic oscillator approximation was implemented within the TI framework. This protocol involved conducting geometry optimizations for both the initial and the final electronic states, followed by vibrational frequency calculations at the optimized geometries. For a general molecule, the DR effect\cite{duschinsky_interpretation_1937} was considered by connecting normal coordinate vectors of the initial ($\mathbf{q}'$) and final  ($\mathbf{q}$) electronic states:
\begin{equation}
\mathbf{q}'=\mathbf{J} \mathbf{q} + \mathbf{k}.\label{eq:dus}  
\end{equation}
Here $\mathbf{J}$ is the Duschinsky rotation matrix, and $\mathbf{k}$ indicates the displacement vector between the PESs of both electronic states along the normal coordinate. $\mathbf{J}$  and $\mathbf{k}$ are given by,
\begin{eqnarray}
    \mathbf{J} &=& (\mathbf{L}')^T\mathbf{L}, \label{eq:J} \\
    \mathbf{k} &=& (\mathbf{L}')^T\mathbf{M}^{1/2}\Delta \mathbf{x}.\label{eq:k}
\end{eqnarray}
Here $\mathbf{M}$ denotes the diagonal matrix of atomic masses, and $\Delta \mathbf{x} \equiv \mathbf{x}-\mathbf{x}'$ is the change between the equilibrium geometries of both states. For a diatomic system, since there is one vibrational mode, matrices $\mathbf{L}$ and $\mathbf{L}'$ become column vectors, and matrix $\mathbf{J}$ in Eqs. (\ref{eq:dus})-(\ref{eq:J}) decays into a scalar ($J=1$), and vectors $\mathbf{q}$, $\mathbf{q}'$, and $\mathbf{k}$ become scalars ($q$, $q'$, and $k$). The FC amplitude for 0-0 transition, $\langle 0|0\rangle$, can then be computed based on $k$ and the vibrational frequencies $\omega'$ and $\omega$.\cite{sharp_franckcondon_1964, ruhoff_recursion_1994, ruhoff_algorithms_2000}  Then, FC amplitudes $\langle 0|n\rangle$ are computed in a recursive manner starting from $\langle 0|0\rangle$.\cite{sharp_franckcondon_1964, ruhoff_recursion_1994, ruhoff_algorithms_2000} The FC factor (FCF) is calculated as the square of the FC amplitude,
\begin{equation}
F_n = \langle 0 |n \rangle^2.
\end{equation}

Meanwhile, the vertical ionization potential in XPS, $I^{\text{vert}}$, and the vertical excitation energy in XAS,  $X^{\text{vert}}$, are respectively calculated by using the $\Delta$Kohn-Sham ($\Delta$KS)  scheme,\cite{PhysRev.139.A619, triguero_separate_1999}
 \begin{eqnarray}
I^{\text{vert}} &=& \it E_{\rm{FCH}}|_{\mathbf{min\ GS}} - E_{\rm{GS}}|_{\mathbf{min\ GS}} + \delta_{\rm{rel}},\\
X^{\text{vert}} &=& \it E_{\rm{XCH}}|_{\mathbf{min\ GS}} - E_{\rm{GS}}|_{\mathbf{min\ GS}} + \delta_{\rm{rel}}.
 \end{eqnarray}
The corresponding adiabatic values are
 \begin{eqnarray}
I^{\text{ad}}  &=& \it E_{\rm{FCH}}|_{\mathbf{min\ FCH}} - E_{\rm{GS}}|_{\mathbf{min\ GS}} + \delta_{\rm{rel}},\\
X^{\text{ad}} &=& \it E_{\rm{XCH}}|_{\mathbf{min\ XCH}} - E_{\rm{GS}}|_{\mathbf{min\ GS}} + \delta_{\rm{rel}},
 \end{eqnarray}
 which consider both electronic and geometrical relaxations. In the above expressions, $E_{\rm{GS}}$, $E_{\rm{FCH}}$, and $E_{\rm{XCH}}$  stand for total energies of the ground electronic state, the  FCH state with one 1s electron removed, and the XCH state with one 1s electron excited to the lowest unoccupied molecular orbital (LUMO), respectively. $\mathbf{min\ GS}$, $\mathbf{min\ FCH}$, or $\mathbf{min\ XCH}$ denotes the optimized structure of each state. $\delta_{\rm{rel}}$ is a small uniform shift considering the differential relativistic effect, which is related to the removal of an electron from the core orbital. For the C, N, and O 1s core holes, $\delta_{\rm{rel}}=$ 0.2, 0.3, and 0.4 eV, were respectively set.\cite{triguero_separate_1999} Based on Eqs. (\ref{eq:ad:xps})-(\ref{eq:ad:xas}) and the vibrational frequencies, the 0-0 vibrational transition energies in XPS and XAS are respectively given by,\cite{hua_theoretical_2020}
  \begin{eqnarray}
I_\text{00}  &=& I^{\text{ad}} +  (\epsilon^e_0 -\varepsilon^g_0) \nonumber\\
&=&I^{\text{ad}} +  \frac{1}{2}(\omega-\omega'),\label{eq:ad:xps} \\
X_\text{00}  &=&X^{\text{ad}} + (\epsilon^e_0 -\varepsilon^g_0)\nonumber\\
&=& 
X^{\text{ad}} + \frac{1}{2}(\omega-\omega').\label{eq:ad:xas}
 \end{eqnarray}
In the above two expressions, the rightmost term stands for the difference in the zero-point vibrational energy (ZPE) of the two electronic states. Hence, the 0-$n$ vibrational ionization and excitation energies are respectively given by,\cite{hua_theoretical_2020}
  \begin{eqnarray}
I_{0n}  &=& I_{00} +n\omega, \\
X_{0n}  &=& X_{00} +n\omega.
 \end{eqnarray}

To broaden the computed FC factors in XPS/XAS, a Lorentzian convolution with respect to the photon/binding energy $E$, with parameters $\Omega$ (center of the Lorentzian) and $\gamma$ (half-width-at-half-maximum),  
 \begin{equation} 
 \Delta(E; \Omega, \gamma) =\frac{1}{\pi} \frac{\gamma}{(E-\Omega})^{2}+\gamma^{2}, 
\label{eq:lorentz}
\end{equation}
was applied. This convolution allowed for the calculation of vibrationally-resolved XPS/XAS spectral intensities:\cite{hua_theoretical_2020}
\begin{eqnarray}
\sigma_{\mathrm{XPS}}\left(E \right) &=& D_{ge}^2 \sum_{n}
\Delta\left(E; I_{0n}, \gamma\right)
\langle 0 | n\rangle^{2}, \label{eq:xps:ti}\\
\sigma_{\mathrm{XAS}}\left(E \right) &=& f_{ge} \sum_{n}
\Delta\left(E ; X_{0n}, \gamma\right)
\langle 0 | n\rangle^{2}.\label{eq:xas:ti}
\end{eqnarray}
In Eq. (\ref{eq:xps:ti}), $D_{ge}$ denotes the electronic transition dipole moments, and we assume it as a constant ($|D_{ge}|$=1) in the calculations. This work considers a simple diatomic case with only one non-equivalent core center, thus it is obvious that this approximation  does not influence the result. We should strengthen that even in a polyatomic molecule, this is still a very good approximation, as verified by the excellent agreement with the experiment in the total C1s XPS of pyridine, contributed by weighted summations from three different types of carbon centers (C$_\alpha$, C$_\beta$, and C$_\gamma$).\cite{hua_theoretical_2020} Similarly, since we consider only one excited state for each system in this work, the oscillator strength $f_{ge}$ in Eq. (\ref{eq:xas:ti}) is taken as a constant term ($f_{ge}$=1) in practical calculations.

 \subsection{Anharmonic approach}

 The other computational approach is to scan the  PECs first, and then to compute the vibrationally-resolved spectra combined with time-dependent wavepacket simulations. Within this approach, the anharmonic effect is considered as embedded in the PECs. By assuming that the photon absorption process is sudden, the initial ($t=0$) wavepacket of the exited state is just the ground state ($n'=0$) vibrational wavefunction of electronic state $g$, i.e., 
 \begin{equation}
 |\Phi_e (t=0)\rangle=|\varphi^{g}_\text{0}\rangle.
 \end{equation}
The time evolution of this wavepacket can be considered as free propagation of  $|\varphi^{g}_\text{0}\rangle$, i.e.,
\begin{eqnarray}
|\Phi_e\it{(t)}\rangle &=& \text{e}^{-i\hat{H}_et}|\varphi^{g}_\text{0}\rangle \nonumber\\ 
&=&\sum_n c_n \exp(-i\epsilon^{e}_nt)|\phi^e_{n}\rangle. 
 \end{eqnarray}
Here $\hat{H}_e$ is the nuclear Hamiltonian of the excited state $e$ and $c_n$ stands for the expansion coefficients. By numerically solving the steady-state Schr\"{o}dinger equation of both states, the nuclear wavefunctions ($|\varphi^{g}_\text{0}\rangle$ and $|\phi^e_{n}\rangle$) and corresponding eigenvalues ($\varepsilon^g_{n'}$ and $\epsilon^{e}_n$) can be obtained.  Then, the time-dependent overlap integral  with the initial vibrational wavefunction (i.e., the auto-correlation function) is given by,
\begin{equation}
S_e (t) =\langle \varphi^{g}_\text{0}|\Phi_e\it{(t)}\rangle.
 \end{equation}

The Lorentzian centered at $\Omega$ in Eq. (\ref{eq:lorentz}) can be rewritten in the time domain as,
\begin{eqnarray}
\Delta(E; \Omega, \gamma) 
&=& - \frac{1}{\pi} \Re \frac{1}{i(E-\Omega)-\gamma} \nonumber\\
&=&-\frac{1}{\pi} \Re \int_0^\infty  \text{exp}\left[i(E-\Omega)t-\gamma t\right]dt,
\end{eqnarray}
where the half Fourier transform was utilized and $\Re$ denotes the real part. The X-ray absorption cross section at incident photon energy $E$ can be calculated as:\cite{couto_carbon_2020, vaz_da_cruz_study_2017}
\begin{equation}
\sigma_\text{{XAS}}\it{(E)} =-\frac{f_{ge}}{\pi}\Re \int_\text{0}^{\infty} S_e(t) \text{exp}\left[i(E-X^\text{res})t-\gamma t\right]dt 
,\label{eq:XAStime}
 \end{equation}
with the resonant energy term given by,
\begin{equation}
    X^\text{res}=X^\text{vert}-\varepsilon^{g}_0.
\end{equation}
Similarly, the intensity of vibrationally-resolved XPS spectroscopy  is described by:
\begin{equation}
\sigma_\text{{XPS}}\it{(E)} =-\frac{D_{ge}^2}{\pi}\Re \int_\text{0}^{\infty} dt \,S_e(t) \text{exp}\left[i(E-I^\text{res})t-\gamma t\right]
,\label{eq:XPStime}
 \end{equation}
with 
\begin{equation}
    I^\text{res}=I^\text{vert}-\varepsilon^{g}_0.
\end{equation}
In practical calculations, the pre-factors 
in Eqs. (\ref{eq:XAStime}) and  (\ref{eq:XPStime}) are taken as constants ($f_{ge}$=1, $|D_{ge}|$=1).

 \section{Computational details} \label{sec:details}
\subsection{Harmonic simulations}
\subsubsection{Electronic structures}

All electronic structure calculations were carried out at the DFT level by using the GAMESS-US package,\cite{gordon_advances_2005, *schmidt_general_1993} enforcing a C$_\text{4v}$ point group symmetry. Four different functionals were chosen: BLYP,\cite{becke_density-functional_1988, lee_development_1988} BP86,\cite{becke_density-functional_1988, perdew_density-functional_1986} B3LYP,\cite{becke_densityfunctional_1993, lee_development_1988} and M06-2X.\cite{zhao_m06_2008} A double basis set technique was used.\cite{hua_theoretical_2020} The aug-cc-pVTZ basis set\cite{dunning_gaussian_1989, *kendall_electron_1992} was used in the geometrical optimization of the ground state. Concerning the excited state, the IGLO-III basis set\cite{diehl_iglo-method_1990} was set for the excited atom. The basis set for the other atom was chosen as follows: in a homonuclear molecule or ion (N$_2$, N$_2^+$), model core potential (MCP) together with corresponding MCP/TZP basis set\cite{sakai_model_1997, *noro_contracted_1997, *bsjp} was employed; while in a heteronuclear system (CO, CO$^+$, NO, NO$^+$), the aug-cc-pVTZ basis was set. Unrestricted DFT (UDFT) was employed for all excited-state calculations except for 1s ionized states of CO, where UDFT encountered SCF convergence errors, and restricted open-shell DFT (RO-DFT) was used instead.  Vibrational frequency calculations were then performed at the optimized structures (all Cartesian coordinates provided in the Supplementary Material\cite{si_zhanglu1}).

\subsubsection{Spectral calculations}
Then, a modified DynaVib package \cite{hua_theoretical_2020} was employed to compute the FCFs based on the obtained energies, geometries, and vibrational frequencies. The vibrationally-resolved XAS and XPS spectra were obtained by convoluting the stick FCFs with a Lorentzian function. A hwhm value, $\gamma$=0.05 eV, was chosen to achieve better agreement with experimental observations. Here $\gamma$ is to phenomenologically include many effects for broadening (such as the lifetime, instrumental, and Doppler broadening, and environmental effects) and similar values have been utilized in previous studies for other systems.\cite{hua_theoretical_2020, wei_vibronic_2022, wei_vibronic_2023, wei_vibronic_2023_iii, cheng_vibrationally-resolved_2022} In principle, one may always use the Voigt convolution (see, e.g., Ref. [\citenum{couto_breaking_2021}]1) to tune a better agreement with the experiment, with the Lorentzian/Gaussian hwhm components read either from a library,\cite{zschornack_handbook_2007} or directly from the experiment.\cite{couto_breaking_2021} Nevertheless, in this work, the hwhm parameter is not a pivotal factor. For simplicity and consistency, a constant hwhm value was utilized in all calculations, which does not impact the validity of our discussions.

\subsection{Anharmonic simulations}
\subsubsection{Potential energy curves by DFT}

The potential energy curves for all four systems (CO$^+$, NO, CO, NO$^+$) were generated using density functional theory (DFT) implemented in the GAMESS-US package.\cite{gordon_advances_2005, *schmidt_general_1993} For CO$^+$ and NO, the lowest 1s excited states of both nuclei are considered, while for CO and NO$^+$, only the lowest O1s excited states are included. The ground state is included for each system. Four different functionals were employed: BLYP, BP86, B3LYP, and M06-2X. The IGLO-III basis set was set for the excited atom and the aug-cc-pVTZ basis set for the other atom. The PEC scan ranges from 0.800 to around 1.600 {\AA} with an incremental of 0.020 {\AA}. In generating the final-state PEC, the converged molecular orbital (MO) from the previous geometry served as the initial guess for the consequent step, ensuring a smooth PEC. In practice, more points were generated at larger distances beyond a certain threshold, determined by observing nonphysical discontinuities indicative of DFT's limitations at large distances. The resulting PECs were fitted with Morse potentials and extrapolated up to 4.000 {\AA}.

\subsubsection{Potential energy curves of CO$^+$ by RASSCF} 

For a selected system CO$^+$, PECs were generated at the RASSCF level by using the Molpro package.\cite{werner_molpro_2012} These include the ground, the lowest C1s and O1s excited states. The C$_\text{2v}$  point group symmetry was enforced at each point. The aug-cc-pVTZ basis set was employed.  In the electronic structure calculations for each snapshot, the state-averaged CASSCF method was employed to calculate the ground-state PEC. The active space consisted of  9 electrons in 7 orbitals, including 2 $\sigma$, 1 $\sigma^*$, 2 $\pi$, and 2 $\pi^*$ orbitals. Three low-lying valence-excited states each were considered for state averaging in the $a_1$, $b_1$, and $b_2$  irreducible representations. Then, the C/O 1s core-excited state PECs were simulated using the RASSCF(11, 1/7/0) method, where the numbers in parentheses represent the total number of active electrons and the size of the RAS1/RAS2/RAS3 spaces. In order to obtain an energy range suitable for spectral analysis, 15 states were averaged for each symmetry ($a_1$, $b_1$, and $b_2$) at both the C and O K-edges. The core orbital was always frozen in the RAS1 space, with an electron occupation number fixed at 1. These procedures have been previously adopted and validated for generating accurate X-ray absorption spectra.\cite{hua_transient_2019, hua_study_2016, zhangyu_nonlinear_2016}

\subsubsection{Wavepacket simulations}
Taking the PECs generated above as inputs, vibrationally-resolved XAS spectra of the four diatomic systems (CO$^+$, NO, CO, NO$^+$) were computed by wavepacket simulations using our in-house XSpecTime program.\cite{XSpecTime} Besides, we also recaptured PEC data from the literature, including PECs of CO$^+$ generated at the RASPT2 level by Couto et al.\cite{couto_carbon_2020} (ground and the lowest C1s and O1s excited states), and PECs of CO generated at the third-order multireference perturbation theory (MRPT3) level by Huang et al.\cite{huang_theoretical_2022} (ground and the lowest O1s excited states). Our tests have shown that the spectra calculated by our code closely match those reported previously,\cite{couto_carbon_2020, huang_theoretical_2022} validating our wavepacket code and providing more extensive comparisons.

The program first reads two sets of PECs, corresponding to the initial and final electronic states, generated by the same electronic structure method (DFT, RASSCF, RAPT2, or MRPT3). Then the spectrum was calculated at 201 discrete points ranging from 0.800 to 4.000 \AA.  The wavepacket propagation was performed with a time step of 1.0 a.u., a total simulation time of 6$\times10^6$ a.u, and a core hole lifetime of 0.05 eV.  It is worth noting that the chosen constant lifetime is the same as that used in the HO study, allowing for a better comparison between the AH and HO simulations. Furthermore, validations have been conducted using shorter time steps and longer simulation times, and the resulting spectra remain unchanged.

\section{Results}
\label{sec:result}

\subsection{Results by the harmonic approach}

\subsubsection{Bond lengths and core-hole induced changes}
Table \ref{tab:bond}  displays the theoretical bond lengths for each system in the optimized ground and excited states and the change induced by core excitation/ionization,
\begin{equation}
\Delta R \equiv R-R'.
\end{equation}

\begin{figure*}
\includegraphics[width=1.0\textwidth]{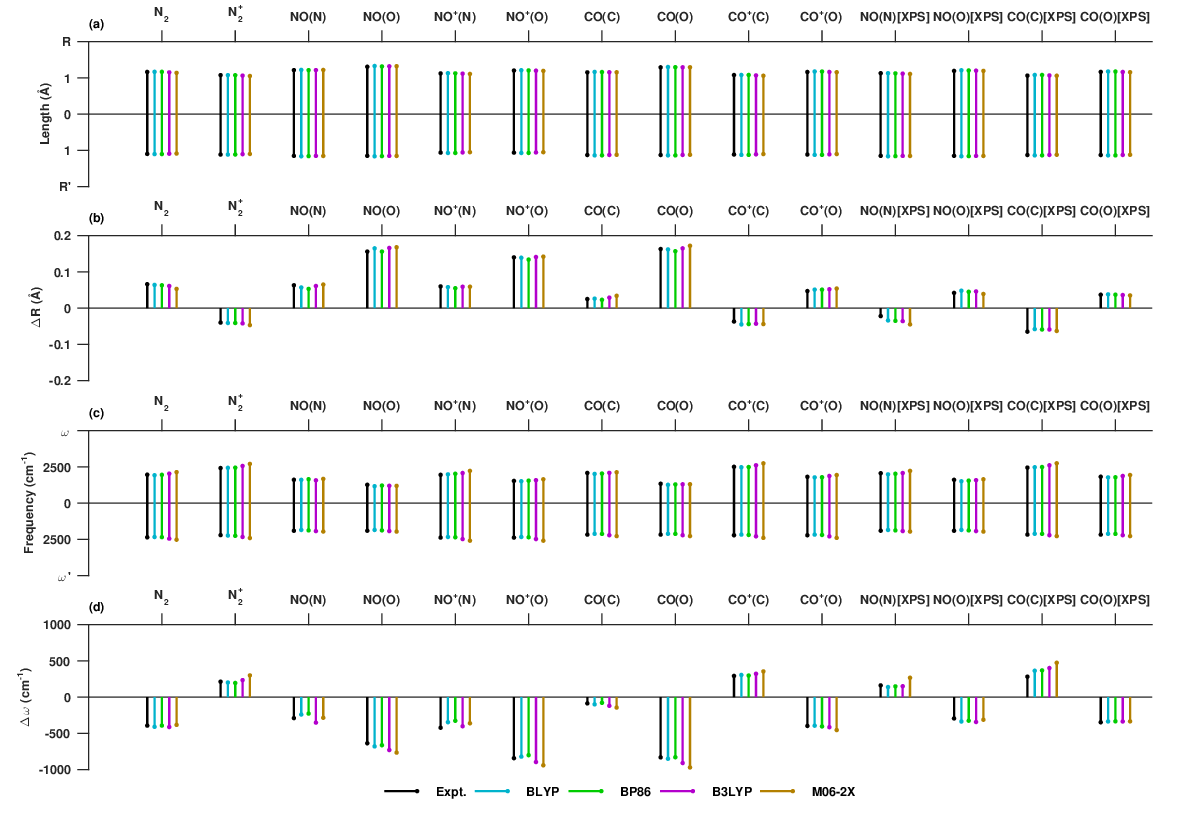}
\caption{Bond lengths  ($R'$, $R$, $\Delta R$) and vibrational frequencies ($\omega'$, $\omega$, $\Delta  \omega$)  of diatomic systems in the ground and the lowest 1s ionized (in XPS; indicated in brackets) or excited (in XAS; the rest) states. DFT with four functionals was used within the harmonic oscillator approximation. (a) $R'$ (ground) and $R$ (excited); (b) $\Delta R=R-R'$; (c) $\omega'$ (ground) and $\omega$ (excited); (d) $\Delta \omega=\omega-\omega'$. Where necessary, the excited atom is specified in parentheses.  Experimental values are recaptured for comparisons. Theoretical and experimental values are listed in Tables~\ref{tab:bond} and \ref{tab:frequency}. Large structural and frequency changes are observed in the lowest O1s excitations in NO, NO$^+$, and CO.}\label{fig:table1}
\end{figure*}

Gas phase experimental data are also included to compare the performance of four functionals. Figure \ref{fig:table1} provides a visual representation of these comparisons.
In the ground state, the experimental bond lengths for these systems range from $R'=$1.063--1.151 {\AA}, with a difference of 0.088 \AA. In the excited states, the bond lengths cover a wider range of 0.244 {\AA} ($R=$1.063--1.307 {\AA}).  The 1s core excitation/ionization can lead to both increases (by 0.025--0.163 \AA) or decreases (by 0.022--0.065 \AA) in bond lengths, depending on the system and the excited atom [indicated by positive or negative  $\Delta R$ in Fig. \ref{fig:table1}(b)]. 

Overall, our theoretical results obtained from all functionals show good agreement with the experiments [Table \ref{tab:bond}; Fig. \ref{fig:table1}(a)]. The mean absolute deviations (MADs) for $R$ ($R'$) predicted by BLYP, BP86, B3LYP, and M06-2X functionals are 0.009, 0.006, 0.006, and 0.011 {\AA} (0.009, 0.008, 0.004, and 0.008 {\AA}), respectively. The maximum absolute deviations (MAXs) are 0.021, 0.017, 0.013, and 0.025  {\AA} (0.012, 0.010, 0.011, and 0.017 {\AA}).  The corresponding relative errors for $R$ ($R'$) are 0.0\%--1.6\%, 0.1\%--1.6\%, 0.1\%--1.2\%, and 0.0\%--2.2\% (0.0\%--1.0\%, 0.1\%--0.9\%, 0.2\%--1.0\%, and 0.3\%--1.5\%).  Regarding $\Delta R$, the deviations from experiments are 0.001--0.012, 0.000--0.013, 0.001--0.014, and 0.001--0.023 {\AA}, respectively. These deviations from experiments are almost imperceptible for $R$ or $R'$ [Fig. \ref{fig:table1}(a)] and small for $\Delta R$  [Fig. \ref{fig:table1}(b)]. 

\subsubsection{Vibrational frequencies and core-hole induced changes}\label{sec:omega}
Table \ref{tab:frequency} presents the vibrational frequencies in the initial ($\omega'$) and final ($\omega$) states, along with comparisons to experiments. The experimental frequencies span a range of 1902.7--2376.7 cm$^{-1}$ (with a difference of 474.0 cm$^{-1}$) in the GS  geometry, and a  wider range of 1266.3--2507.6 cm$^{-1}$ (with a difference of 1241.3 cm$^{-1}$) in the excited state, corresponding to typical frequencies for stretching vibrational modes. The mean absolute deviations for BLYP, BP86, B3LYP, and M06-2X are 47.7, 37.3, 68.1, and 145.5 {cm$^{-1}$} for $\omega$, and 47.2, 31.6, 59.2, and 127.6 {cm$^{-1}$} for  $\omega'$. The maximum deviations are 97.7, 73.0, 160.6, and 296.3 {cm$^{-1}$} for $\omega$ (and 55.3, 48.8, 122.8, and 210.9 for  $\omega'$).  The relative deviations for $\omega$ ($\omega'$) are 0.3\%--7.7\%, 0.6\%--4.0\%, 0.4\%--6.6\%, and 2.3\%--13.9\% (1.0\%--2.9\%, 0.6\%--2.3\%, 1.3\%--5.6\%, and 3.0\%--9.1\%).  

Table \ref{tab:frequency} also shows the vibrational frequency change between the initial and final states, denoted by, 
\begin{equation}
\Delta\omega \equiv   \omega - \omega'.
\end{equation}
Correspondingly, the mean absolute deviations in $\Delta\omega$ are in the range of 3.5--82.0, 0.8--96.2, 11.4--119.5, and 4.7--192.3 {cm$^{-1}$}. Similarly to the bond lengths, the core-hole-induced frequency change does not follow a definite trend [Fig. \ref{fig:table1}(d)]. Deviations between our DFT calculations to experiments appear to be larger for frequencies compared to bond lengths. This is because vibrational frequency corresponds to the second-order energy derivatives of the Cartesian coordinates. The deviations for $\omega'$ or $\omega$ are small for BLYP, BP86, and B3LYP, and acceptable for M06-2X, as shown in Fig. \ref{fig:table1}(c).  In terms of  $\Delta\omega$, the deviations by the former three are also relatively small, while results by M06-2X are acceptable for most examples, as displayed in Fig. \ref{fig:table1}(d). The largest deviation occurs in C1s XPS of CO, with a value of 192.3 {cm$^{-1}$} deviation by M06-2X. 

The analysis of vibrational frequencies supports the same conclusion as the bond lengths: the BLYP, BP86, and B3LYP functionals yield better results compared to M06-2X.  Our DFT calculations demonstrate reasonable accuracy in predicting vibrational frequencies, considering the absolute values of experimental frequencies are some 1900-2400 (1200-2500) cm$^{-1}$ in the ground (excited) state, and show balanced accuracy in both states, as indicated by the similar MAD values for the same functional. In other studies, similar levels of accuracy have been reported. For instance, Moitra et al.\cite{moitra_vibrationally_2020} achieved an overestimation of 90 cm$^{-1}$  (2233 cm$^{-1}$ predicted \textit{versus} 2143 cm$^{-1}$ experimental) for CO  and 102 cm$^{-1}$  for N$_2$ (2432 cm$^{-1}$ predicted \textit{versus} 2330 cm$^{-1}$ experimental) ,  in the ground state by MP2. 

\subsubsection{XAS and XPS of diatomic systems: An  overview}
Figures \ref{fig:xas:n2:n2+}-\ref{fig:xas:co:co+} depict the simulated vibrationally-resolved XAS spectra of N$_2$, N$_2^+$, NO, NO$^+$, CO, and CO$^+$, which were computed using  DFT with four functionals. These spectra will be discussed in detail in subsequent sections. In general, the theoretical results show good agreement with experimental observations for most case studies. However, noticeable differences between different functionals can often be observed, although the level of sensitivity may depend on the specific system and spectroscopy being studied.

\subsubsection{XAS of N$_2$ and N$_2^+$}

\begin{figure*}
\includegraphics[width=0.6\textwidth]{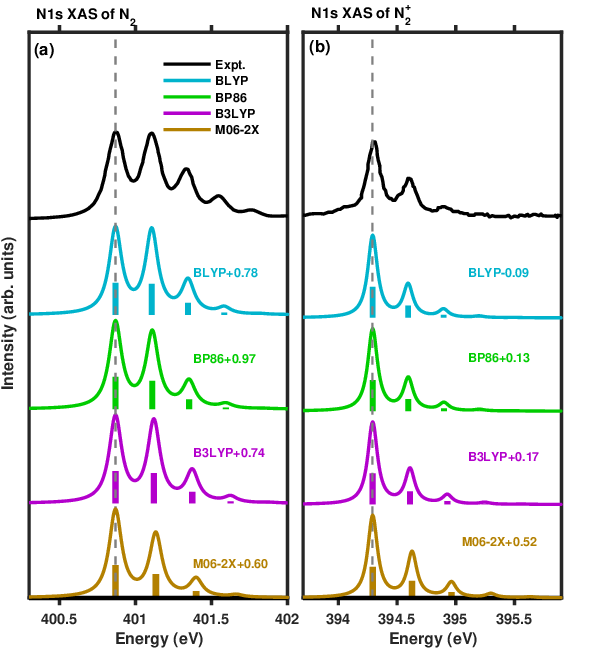}
\caption{Vibrationally-resolved N1s XAS spectra of (a) N$_2$ and (b) N$_2^+$ simulated by  DFT with different functionals by using the harmonic oscillator approximation. To better compare with the experiments of N$_2$\cite{chen_k_1989, hitchcock_k-shell_1980} and N$_2^+$,\cite{lindblad_x-ray_2020} theoretical spectra have been shifted (indicated by numbers in eV) by aligning the 0-0 peak (indicated by dashes).
}\label{fig:xas:n2:n2+}
\end{figure*}

Figure \ref{fig:xas:n2:n2+} displays the computed N1s XAS spectra of N$_2$ and N$_2^+$ compared to the experiments.\cite{chen_k_1989, hitchcock_k-shell_1980, lindblad_x-ray_2020} In Fig. \ref{fig:xas:n2:n2+}(a), the experimental spectrum exhibits five distinct vibronic features, with the 0-0 and 0-1 peaks having nearly equal intensities.  All four functionals accurately predict the peak separations (corresponding to the vibrational frequencies) and relative peak intensities.  BLYP and B3LYP yield almost identical intensities for the 0-0 and 0-1 peaks, showing the best agreement with the experiment. BP86 shows a slightly weaker  0-1 peak than the 0-0 peak, and M06-2X shows an even weaker 0-1 peak. M06-2X predicted relatively larger peak separations compared to the experiment. 

In Fig. \ref{fig:xas:n2:n2+}(b), three vibronic features are evident in the experimental spectrum, with the 0-1 peak having approximately half the intensity of the 0-0 peak. All four functionals agree well with the experiment. M06-2X slightly overestimated the peak separations but best reproduced the 0-1 peak intensity as compared to the 0-0 peak.

\subsubsection{XAS of NO and NO$^+$}

Figure \ref{fig:xas:no:no+}(a-b) presents the simulated XAS spectra of  NO  at the N1s and O1s edges. At the N1s edge (panel a), all four functionals align closely with the fitted spectrum to the lowest N1s state ($^2\Delta$) from the experiment.\cite{remmers_high-resolution_1993, wang_filtering_2001} Among the functionals, M06-2X  exhibits a relatively stronger 0-1  peak compared to the others. At the O1s edge (panel b), the available spectrum of the lowest state ($^2\Sigma^-$) fitted of the experiemnt\cite{puttner_vibrationally_1999} has limited resolution, showing a broad peak with several distinguishable weak structures. The predictions of the fine structure using the four functionals were consistent and acceptable, although they tend to predict relatively larger peak separations compared to the experiment and overestimate the peaks in the higher-energy region. This may be attributed to significant changes in the PEC induced by O1s core excitation.\cite{wei_vibronic_2023_iii} Specifically, there is an increase in bond length by +0.156 {\AA} (from 1.151 to 1.307 {\AA}) and a decrease in vibrational frequency by -636.4 cm$^{-1}$ (from 1902.7 to 1266.3 cm$^{-1}$). Among all the examples studied, these changes represent the second-largest change in bond length (Table  \ref{tab:bond}) and the third-largest change in vibrational frequency (Table \ref{tab:frequency}).

Figure \ref{fig:xas:no:no+}(c-d) showcases the XAS results of NO$^+$ at the N1s and O1s edges. At the N1s edge (panel c), all four functionals show good agreement with the experimental spectrum,\cite{lindblad_experimental_2022} which displays five distinct vibronic features with approximately equal intensities for the 0-0 and 0-1 peaks. The pure functionals accurately predict the peak separations, although they slightly underestimate the intensity of the 0-1 peak and exhibit a faster decay with increasing vibrational quantum number $n$. On the other hand, the hybrid functionals (B3LYP and M06-2X) slightly overestimate the peak separations but match the experiment well in terms of peak intensities, with the 0-0 and 0-1 peaks having similar intensities.

In the O1s edge (panel d), the experimental spectrum\cite{lindblad_experimental_2022} displays more intricate structures with ten relatively weak vibronic features. Accurately identifying the 0-0 peak seems somewhat challenging.  All four functionals produce similar theoretical spectra that show acceptable agreement with the experiment.  Although the overall shapes of the peak profiles are consistent, the theoretical predictions underestimate the peaks in the higher-energy region (i.e., those with larger vibrational quantum numbers $n$).

The discrepancy between the theoretical and experimental O1s XAS spectra of NO$^+$ can be attributed to the harmonic oscillator approximation used. A better agreement with the experiment was achieved by considering anharmonic effects through the use of RASPT2 PECs.\cite{couto_breaking_2021} The underlying reason lies in the significant changes in PEC    induced by O1s core excitation. These changes include shifts in equilibrium position and alterations in curvature. As mentioned above, the O1s core excitation in NO$^+$ resulted in an increase of +0.140 {\AA} (from 1.063 to 1.203 \AA) in bond length and a decrease of  -841.8 cm$^{-1}$ (from 2376.7 to 1534.9 cm$^{-1}$) in vibrational frequency. Among all the examples studied, these changes represent the third-largest change in bond length (Table \ref{tab:bond}) and the largest vibrational frequency change (Table \ref{tab:frequency}), respectively. 

\subsubsection{XAS of CO and CO$^+$}
Figure \ref{fig:xas:co:co+}(a-b) shows the XAS spectra of CO at the C1s and O1s edges. CO and NO$^+$ are isoelectronic species, and the results for CO show very similar trends to NO$^+$.  At the C1s edge (panel a), the experimental spectrum exhibits only two simple features, while at the O1s edge (panel b), approximately 12 weak features are observed within a broad peak. All functionals demonstrate weak functional dependence and predict good agreement with the experiment at the C1s edge, and acceptable agreement at the O1s edge.  However, similar to the case of NO$^+$ at the O1s edge, all theories underestimate the intensities of high-energy features.

The underlying structural reason is also the significant changes in the PEC as induced by O1s core excitation. Specifically, the bond length increases by +0.163 {\AA} (from 1.128 to 1.291 \AA) and the vibrational frequency decreases by -830.9 cm$^{-1}$ (from 2169.8 to 1338.9 cm$^{-1}$). Among all examples studied, these changes represent the largest change in bond length (Table \ref{tab:bond}) and the second-largest change in vibrational frequency (Table \ref{tab:frequency}), respectively.

The XAS spectra of CO$^+$ at the C1s and O1s edges are much simpler. Figure \ref{fig:xas:co:co+}(c-d) shows the XAS spectra of CO$^+$ at both edges. All four functionals demonstrate good agreement with the experiment in terms of energies and profiles, exhibiting only weak functional dependence \cite{couto_carbon_2020}.

\subsubsection{XPS of CO and NO}
\begin{figure*}
\includegraphics[width=1.0\textwidth]{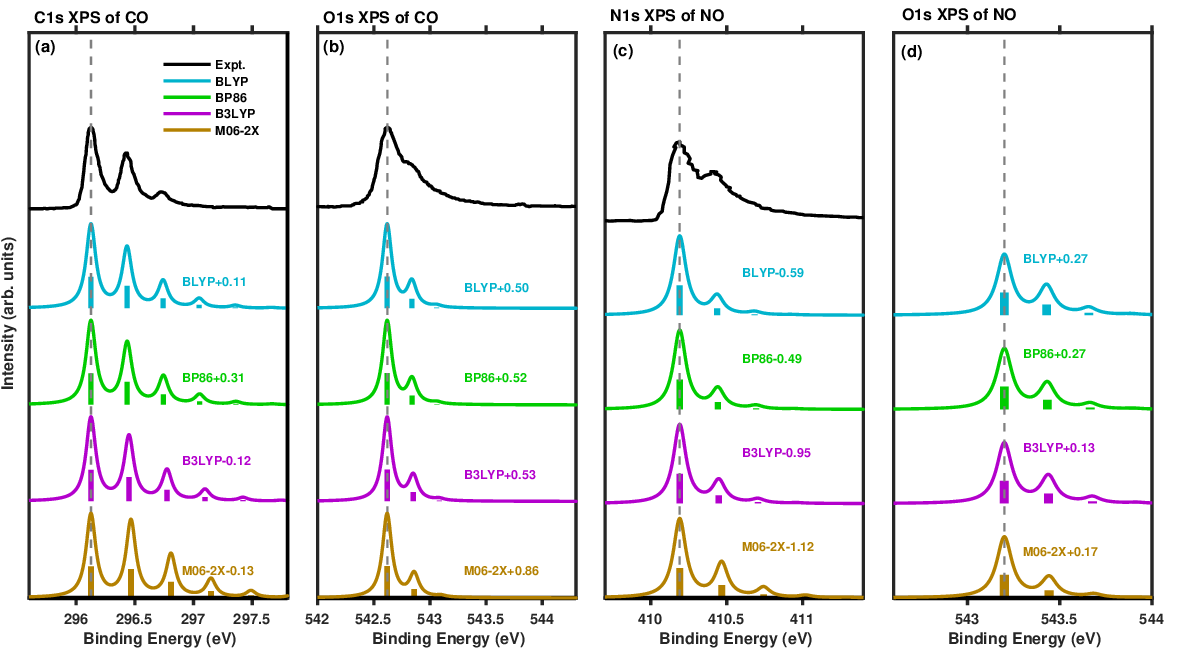}
\caption{Vibrationally-resolved XPS spectra of (a-b) CO  and (c-d) NO  simulated by  DFT with different functionals within the framework of the harmonic oscillator approximation: (a) C1s, (b) O1s, (c) N1s, and (d) O1s edges. To better compare with the experiments of CO (C1s and O1s edges) \cite{hergenhahn_vibrational_2004} and NO (N1s edge)\cite{hoshino_vibrationally_2008, bagus_anomalous_1974}, theoretical spectra have been shifted by aligning the 0-0 peak (dashes). In panel (d), a similar calibration was performed to match the experimental O1s BE of 543.2 eV \cite{bakke_table_1980}. All shifts are indicated by numbers in eV.
}\label{fig:xps:co:no}
\end{figure*}

All examples above are on XAS spectra. The focus now shifts to XPS spectra of CO and NO. Figure \ref{fig:xps:co:no}(a-b) presents the simulated vibrationally-resolved XPS spectra of CO at the C and O 1s edges. In comparison to the XAS experiments, both edges show a blue shift in the 0-0 peak: +8.84 eV (from 287.29 to 296.13 eV)  and +9.00 eV (from 533.62 to 542.62 eV). The experimental XPS spectra \cite{hergenhahn_vibrational_2004} exhibit simple vibronic features at both edges, and all four functionals accurately reproduced the experimental spectra. However, it is worth noting that the hybrid functional M06-2X  slightly overestimated the intensity of the 0-1 peak in the C1s edge [Fig. \ref{fig:xps:co:no}(a)].

Figure \ref{fig:xps:co:no}(c-d) presents the simulated vibrationally-resolved XPS spectra of NO at both the N and O 1s edges. In comparison to the XAS experiments,  both edges show blue shifts in the 0-0 peak: +10.80 eV (from 399.39 to 410.19 eV, see Table \ref{tab:energies})  and +11.72 eV (from 531.48 to 543.20 eV). In Fig. \ref{fig:xps:co:no}(c), the experimental N1s XPS spectrum\cite{hoshino_vibrationally_2008, bagus_anomalous_1974} shows simple vibronic features, resolving clean 0-0 and 0-1 peaks. All four functionals accurately reproduced the experimental spectrum. Figure \ref{fig:xps:co:no}(d) displays only theoretical spectra, since no experimental spectrum is available to our knowledge. Nevertheless, we used the reported experimental BE of 543.20 eV\cite{bakke_table_1980} to calibrate our simulated 0-0 peak to this value, where all four methods exhibit small shift values.

By comparing the XAS [Fig. \ref{fig:xas:co:co+}(b) and Fig. \ref{fig:xas:no:no+}(b)] and XPS [Fig. \ref{fig:xps:co:no}(b) and Fig. \ref{fig:xps:co:no}(d)] spectra of CO and NO at the O1s edge, one can find the different responses of the final-state PEC induced by O1s excitation and ionization. This may be related to the odd or even number of electrons. 

\subsection{Results by the anharmonic approach}


\subsubsection{PECs and spectroscopic constants}

\begin{figure*}
\includegraphics[width=1.0\textwidth]{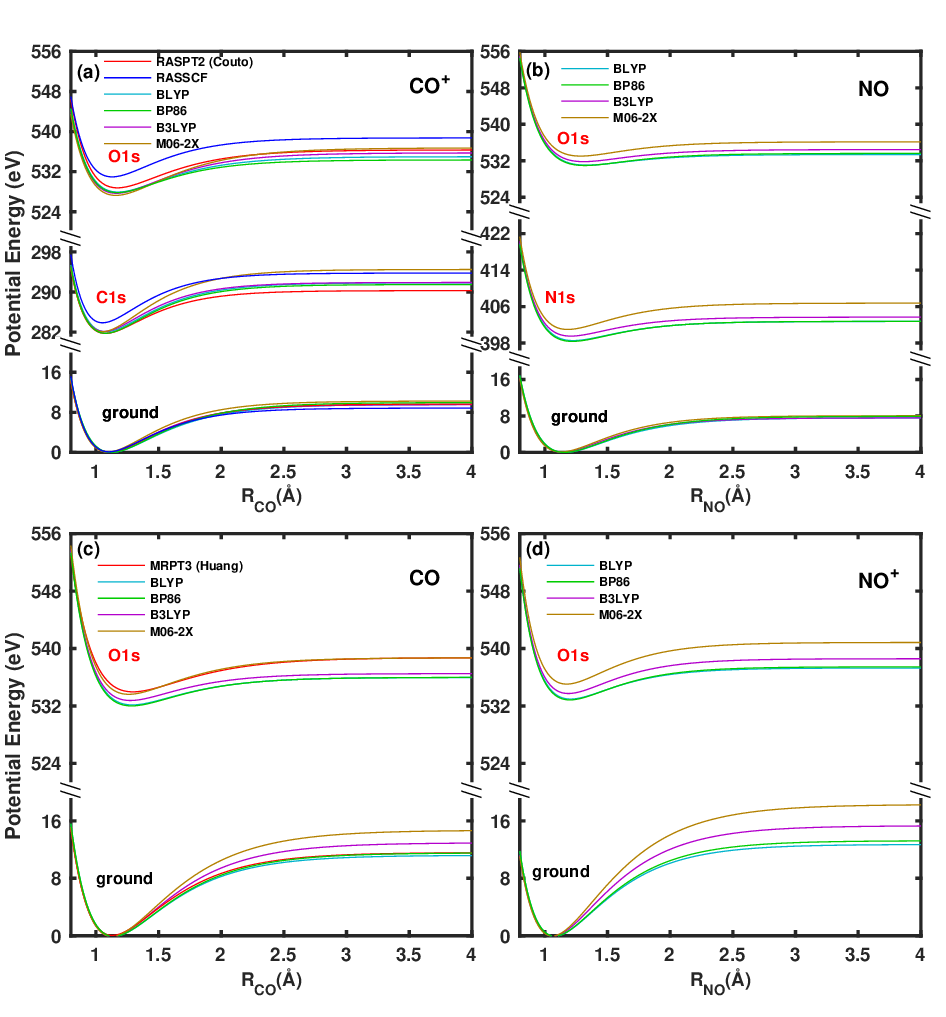}
\caption{Simulated PECs of (a) CO$^+$, (b) NO, (c) CO  and (d) NO$^+$, in the ground and the lowest C/N/O 1s electron excited states by using different theoretical methods. In panel a, RASPT2 curves were taken from Couto et al.\cite{couto_carbon_2020} In panel c, MRPT3 curves of CO  were taken from Huang et al.\cite{huang_theoretical_2022}
}\label{fig:pes} 
\end{figure*}

Figure \ref{fig:pes}(a-d)  shows computed PECs of four selected systems, CO$^+$, NO, CO, and NO$^+$, by using various electronic structure methods.  These PECs are fitted to the Morse potential, and the resulting parameters can be found in Table  \ref{tab:morse}.  For illustration purposes, we will focus on the CO$^+$ system. The PEC curves for the other systems exhibit similar behavior and also fit well to the Morse potential, which will be discussed in the subsequent spectral analysis.

As shown in Fig. \ref{fig:pes}(a), in the ground state, the DFT and multiconfigurational methods show similar profiles, with equilibrium positions ($R_\text{e}$) ranging from 2.066 to 2.121 bohr. The relative errors, compared to the experimental bond length of 2.108 bohr (1.115 \AA)\cite{neeb_coherent_1994}, range from 0.55\% (BP86) to 1.99\% (M06-2X).  These relative deviations are comparable to the discrepancies observed in bond lengths obtained from geometry optimizations, which range from 0.45\% (B3LYP) to 1.17\% (M06-2X).

However, larger discrepancies are observed in PECs of the lowest C1s and O1s excited states of CO$^+$ [Fig. \ref{fig:pes}(a)]. In the C1s state, the equilibrium positions ($R_e$) range from 1.990 to 2.039 bohr, while in the O1s state, they range from 2.134 to 2.222 bohr. The corresponding relative errors vary from 0.20\% (RASPT2) to 1.95\% (RASSCF) for the C1s state, and from 0.04\% (B3LYP) to 2.87\% (RASSCF) for the O1s state, when compared to experimental values.\cite{couto_carbon_2020} These relative errors are consistent with the discrepancies observed in bond lengths obtained from geometry optimizations: 0.19\% to 1.76\% for the C1s state and 0.09\% to 1.20\% for the O1s state.

Regarding the fitted dissociation energy or well depth ($D_\text{e}$) of CO$^+$, in the ground state, it falls within the range of 0.32-0.37 a.u., with  RASSCF yielding the smallest and M06-2X producing the largest value. In the C1s and O1s excited states, $D_\text{e}$ ranges 0.31--0.46 a.u. and 0.24--0.35 a.u., respectively. 

\subsubsection{C1s and O1s XAS of CO$^+$}
Figure \ref{fig:co+:td:xas}(a) depicts the C1s XAS spectra of CO$^+$ computed using the anharmonic method with different PECs. Generally, all AH spectra essentially agree well with the experiment,\cite{couto_carbon_2020}  although they exhibit some noticeable method dependence. The RASPT2 method tends to underestimate the 0-1 peak intensity, while RASSCF overestimates it. Additionally, over a wider energy range, the RASPT2 (RASSCF) spectral intensity decays much faster (slower)  with the increasing vibrational quantum number $n$, compared to the experimental trend. In other words, the experiment spectrum is in between the RASSCF and  RASPT2 predictions. Both methods adequately account for static correlation effects. However, RASSCF only treats dynamic correlation within the active space, whereas RASPT2 includes more dynamic correlation effects by perturbatively treating the effects of inactive orbitals.  These results indicate the sensitivity of these calculations to the dynamic correlation. 

In the case of DFT methods, there is substantially less variability among different functionals, indicating weak functional dependence. For a given functional, HO and AH spectra largely overlap, although some differences are observable. This is particularly evident with the BP86 functional, where AH spectra align excellently with the experiment, while HO spectra noticeably underestimate the 0-1 peak intensity. These discrepancies can be attributed to the anharmonic effects.

Figure \ref{fig:co+:td:xas}(b) demonstrates a similar performance at the O1s edge. In contrast to the C1s edge, RASPT2 (RASSCF) overestimates (underestimates) the 0-1 peak and exhibits slower (faster) decay with $n$, indicating the significant influence of dynamic correlation treatment. At the DFT level, AH models generally predict a slower decay with $n$ compared to their corresponding HO method. The agreement with the experimental results depends on the functional used. For instance, with the BP86 functional, the AH calculation accurately reproduces the experimental results, surpassing the HO method. Conversely, with the M06-2X functional, the HO calculation generates a better fit than its AH counterpart.

\subsubsection{N1s and O1s XAS of NO}
Figure \ref{fig:no:td:xas} compares XAS spectra of NO simulated by harmonic and anharmonic methods by using different DFT functionals. At the N1s edge [Fig. \ref{fig:no:td:xas}(a)], although the influences of functional and anharmonicity are discernible, the spectra generally agree well with the fitted spectrum for the lowest N1s state ($^2\Delta$) of the experiment.\cite{remmers_high-resolution_1993, wang_filtering_2001} M06-2X shows a relatively larger difference between the HO and AH spectra than other methods.

In Fig. \ref{fig:no:td:xas}(b), the O1s XAS spectra of NO display more intricate vibronic structures compared to those at the N1s edge. This complexity arises from substantial geometric alterations induced by the O1s core hole, and as a result, the 0-0 peak is not dominant. For each functional, a notable discrepancy exists between HO and AH methods. This can be attributed to the shallower PECs, which deviate more from harmonic oscillators compared to other molecules. Table \ref{tab:morse} shows that the well depth of the Morse potential, or dissociation energies $D_\text{e}$,  for the NO O1s excited state (0.09--0.11 a.u.) is noticeably smaller than its N1s excited states (0.15--0.21 a.u.) and also smaller than  $D_\text{e}$ values of other molecules (CO O1s, 0.14--0.19 a.u.; NO$^+$ O1s, 0.16--0.21 a.u.;  CO$^+$ O1s, 0.24--0.34; CO$^+$ C1s, 0.31--0.46 a.u.)

\subsubsection{O1s XAS of CO and NO$^+$} \label{sec:td:difficult}
As shown in Fig. \ref{fig:conop:td:xas}(a-b), the O1s XAS spectra of CO  and NO$^+$  exhibit more complex vibronic structures than those observed for CO$^+$ due to larger geometrical changes induced by the O1s core hole. For each functional, a noticeable discrepancy between the AH and HO methods, with AH spectra predicting larger intensities mainly at the higher energy peaks (and sometimes also at the lower energy peaks). The inclusion of anharmonic effects in the AH calculations accounts for this difference, thereby improving upon the HO approach.

Figure \ref{fig:conop:td:xas:max} provides a detailed comparison between the AH results and experimental data. Multiconfigurational methods (MRPT3 or RASPT2) show the best agreement with the experiments. In these spectra, each peak is identified by the vibrational quantum number $n$ of the final state, and the one that labels the peak with the maximum intensity is denoted as $n_\text{max}$. Remarkably, the multiconfigurational methods predict the same $n_\text{max}$ values ($n_\text{max}$=5 for CO; $n_\text{max}$=4 for NO$^+$) as those observed experimentally, while DFT predicts smaller $n_\text{max}$ values (mostly by 1). Our results highlight the strong sensitivity of $n_\text{max}$ to the quality of the PECs.
 
Beyond the $n_\text{max}$ prediction, the BLYP, BP86, and B3LYP functionals show good agreement with experimental results in terms of both peak separations and intensities, despite their ease of implementation. In the CO spectra, BLYP/BP86 exhibits a decay speed relative to $n$ that matches the experimental data, whereas MRPT3 predicts a slower rate of decay. These calculations underscore the solid performance of our DFT method, offering competitive accuracy compared to the more advanced and computationally intensive multiconfigurational methods.

\section{Discussion}
\label{sec:discussion}

\subsection{Time-independent \textit{versus} time-dependent implementations}

In this work, the performance of harmonic and anharmonic approaches was comprehensively compared, and the anharmonic effect was found sensitive in some systems. The two approaches were respectively implemented in the TI and TD frameworks. It is necessary to mention that the choice of framework does not impact the results. In physics and chemistry, many problems can be equivalently addressed either in the frequency or the time domain. 

The combination of the TI solution and the harmonic approximation is widely used in the literature,\cite{cerezo_fcclasses3_2023}, especially for polyatomic molecules. This is because the TI solution allows for explicit spectral interpretations and provides direct insights into structural changes when considering the Duschinsky rotation effect.  In this study, our TI implementation for the harmonic approach was guided by the success of this protocol in many polyatomic systems.\cite{hua_theoretical_2020, wei_vibronic_2022, wei_vibronic_2023, wei_vibronic_2023_iii, cheng_vibrationally-resolved_2022} Meanwhile, the TD implementation would yield consistent results, as verified by tests performed with our DynaVib code.\cite{DynaVib} Although the TD implementation does not provide explicit access to identify individual vibronic transitions from peak features, it offers computational efficiency advantages, particularly when addressing problems with significant temperature effects.\cite{sun_thermally_2019} When studying diatomic molecules, there is more flexibility in selecting different combinations. The TD implementation of the anharmonic approach adopted in this work is commonly utilized in the literature for diatomic studies.\cite{couto_breaking_2021, couto_carbon_2020} Although both TD and TI solutions based on the same input PECs can lead to consistent vibrationally-resolved (stead-state) X-ray spectra, the TD framework can be easily extended to study time-resolved signals.

\subsection{Functional dependence in diatomic \textit{versus}  polyatomic systems}

In our DFT simulations conducted on these diatomic systems,  all four functionals generally produce spectral and structural data in good agreement with experiments. Nevertheless, the functional dependency is still evident, at least more noticiable than in polyatomic molecules. Among the four functionals, M06-2X occasionally performs less effectively than the other three. 

The functional dependence on core binding energies and X-ray spectra is a well-known issue in DFT\cite{du_theoretical_2022, wei_vibronic_2022, norman_simulating_2018, li_first-principles_2012}  as well as TDDFT\cite{annegarn_combining_2022, imamura_description_2007} calculations. Extensive studies have been conducted on  binding energies,\cite{du_theoretical_2022, norman_simulating_2018, pueyo_bellafont_prediction_2015, pueyo_bellafont_performance_2016} and electronic-only\cite{fronzoni_vibrationally_2014, li_first-principles_2012,peverati_m11-l_2012, labat_density_2007, cavigliasso_accurate_1999} and vibrationally-resolved\cite{wei_vibronic_2022, fronzoni_vibrationally_2014} X-ray spectra,  particularly for polyatomic systems. It is known that the degree of functional dependence varies across different systems.\cite{norman_simulating_2018} For instance, in the polyatomic molecule pyrimidine (C$_4$H$_4$N$_2$), weak functional dependence was observed in the vibrationally-resolved N1s XPS spectrum.\cite{wei_vibronic_2022} The functional sensitivity appears to be diminished in polyatomic systems due to the presence of multiple vibrational modes. That is, the combined effects of various vibrational modes tend to wash out the functional dependence. 

\subsection{Molecule \textit{versus} cation}

Here we concentrate on examining only the XAS outcomes of neutral molecules and their corresponding cations. By examining the spectra and energies summarized in Table \ref{tab:energies}, we can explore the influence of charge state or parity (odd or even number of electrons) on various spectra.

\begin{figure*}
\includegraphics[width=1.0\textwidth]{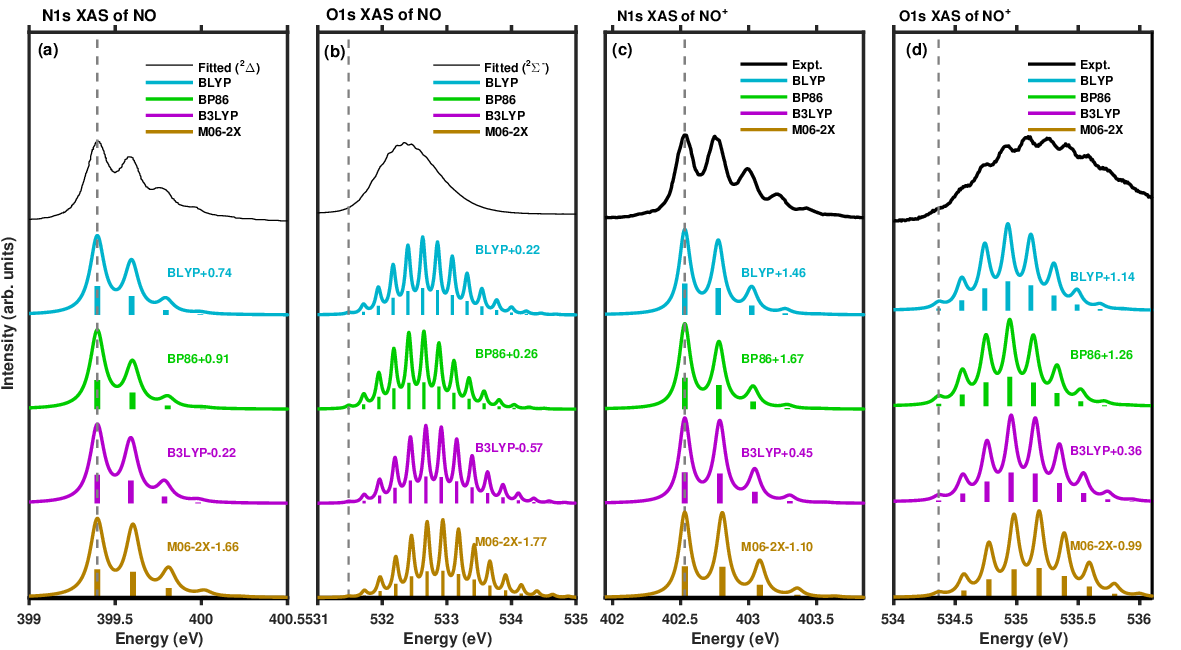}
\caption{Vibrationally-resolved XAS spectra of (a-b) NO and (c-d) NO$^+$ simulated by DFT with different functionals by using the harmonic oscillator approximation: (a,c) N1s and (b,d) O1s edges. To better compare with the fitted spectra for the lowest N1s ($^2\Delta$)\cite{wang_filtering_2001, remmers_high-resolution_1993} and O1s ($^2\Sigma^-$)\cite{puttner_vibrationally_1999} states of the experiments, theoretical spectra have been shifted (indicated by numbers in eV) by aligning the first peak. To better compare with the experiment\cite{lindblad_experimental_2022} of NO$^+$, theoretical spectra have been shifted by aligning the 0-0 peak (dashes). 
}\label{fig:xas:no:no+}
\end{figure*}

\begin{figure*}
\includegraphics[width=1.0\textwidth]{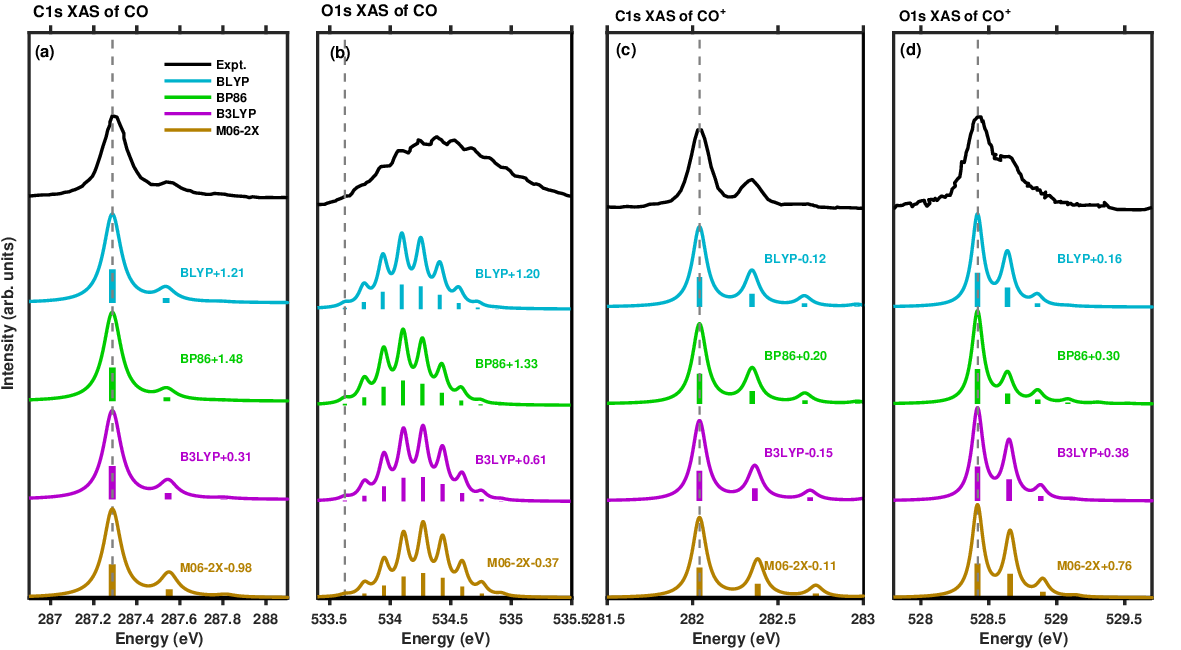}
\caption{Vibrationally-resolved XAS spectra of (a-b) CO and (c-d) CO$^+$ simulated by  DFT with different functionals by using the harmonic oscillator approximation: (a,c) C1s and (b,d) O1s edges. To better compare with the CO (C1s\cite{hitchcock_k-shell_1980} and O1s\cite{puttner_vibrationally_1999} edges) and  CO$^+$\cite{couto_carbon_2020} experiments, theoretical spectra have been shifted (indicated by numbers in eV) by aligning the 0-0 peak (dashes).
}\label{fig:xas:co:co+}
\end{figure*}

\begin{figure*}
\includegraphics[width=0.6\textwidth]{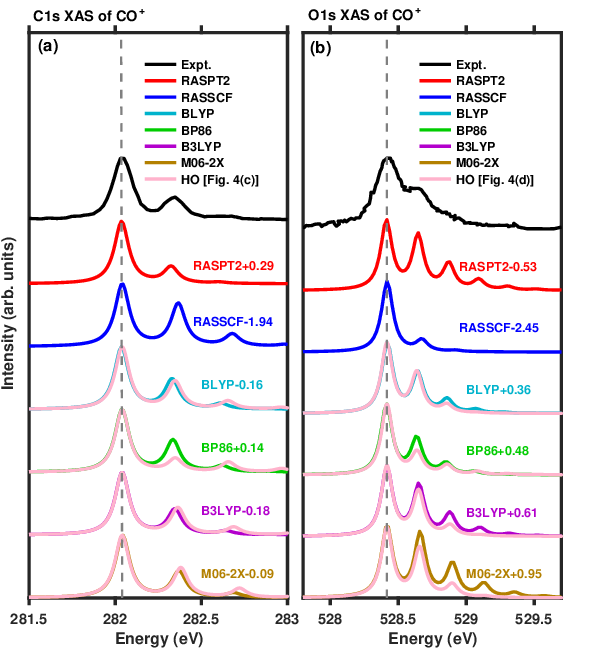}
\caption{Comparison of vibrationally-resolved XAS of CO$^+$ simulated by the anharmonic (other colored lines) and harmonic [pink lines; recaptured from Fig. \ref{fig:xas:co:co+}(c-d)] methods: (a) C1s and (b) O1s edges. Different electronic structure methods are used, where ``RASPT2'' denotes our calculated spectra using RASPT2 PECs [see Fig. \ref{fig:pes}(a)] from Couto et al,\cite{couto_carbon_2020} while the rest of theoretical lines are purely our results. To better compare with the experiments,\cite{couto_carbon_2020} theoretical spectra have been shifted (indicated by numbers in eV) by aligning the 0-0 peak (dashes). } 
\label{fig:co+:td:xas}
\end{figure*}

\begin{figure*}
\includegraphics[width=0.6\textwidth]{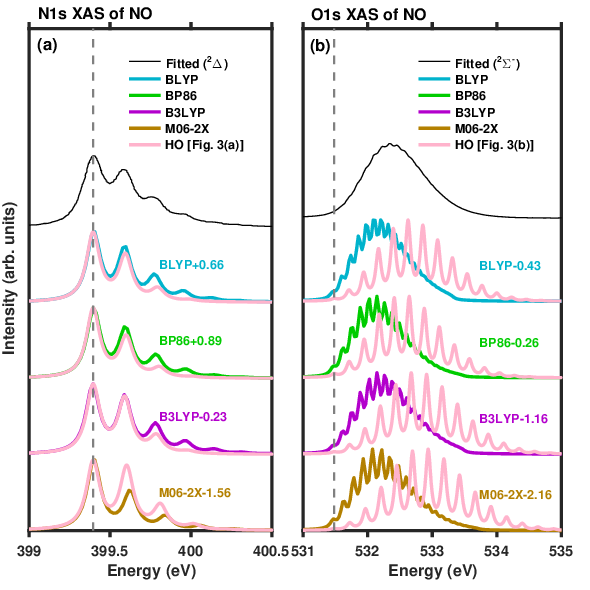}
\caption{Comparison of vibrationally-resolved (a) N1s and (b) O1s XAS spectra of NO simulated by the anharmonic (other colored lines) and harmonic [pink lines; recaptured from Fig. \ref{fig:xas:no:no+}(a-b)] methods. Different functionals are used. To better compare with the fitted spectra for the lowest N1s ($^2\Delta$)\cite{wang_filtering_2001, remmers_high-resolution_1993} and O1s ($^2\Sigma^-$)\cite{puttner_vibrationally_1999} states of the experiments, theoretical spectra have been shifted (indicated by numbers in eV) by aligning the 0-0 peak (dashes). 
 }
\label{fig:no:td:xas}
\end{figure*}

\begin{figure*}
\includegraphics[width=0.6\textwidth]{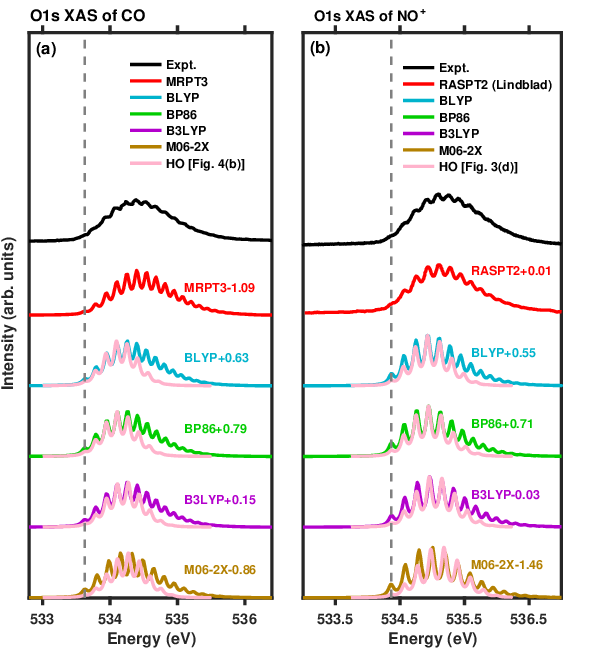}
\caption{Comparison of vibrationally-resolved XAS spectra of (a) CO and (b) NO$^+$   simulated by the anharmonic (other colored lines) and harmonic [pink lines;  recaptured from Figs.  \ref{fig:xas:co:co+}(b) and \ref{fig:xas:no:no+}(d), respectively] methods. Different electronic structure methods are used. In panel a, ``MRPT3''  denotes our calculated spectrum using MRPT3 PECs [see Fig. \ref{fig:pes}(c)] from Huang et al.\cite{huang_theoretical_2022}  In panel b, the RASPT2 spectrum was taken from Lindblad et al.\cite{lindblad_experimental_2022} The rest of theoretical lines are purely our results. To better compare with the experiments of CO\cite{puttner_vibrationally_1999} and NO$^+$,\cite{lindblad_experimental_2022}   theoretical spectra have been shifted (indicated by numbers in eV) by aligning the first peak (dashes).
} 
\label{fig:conop:td:xas}
\end{figure*}

\begin{figure*}
\includegraphics[width=1.0\textwidth]{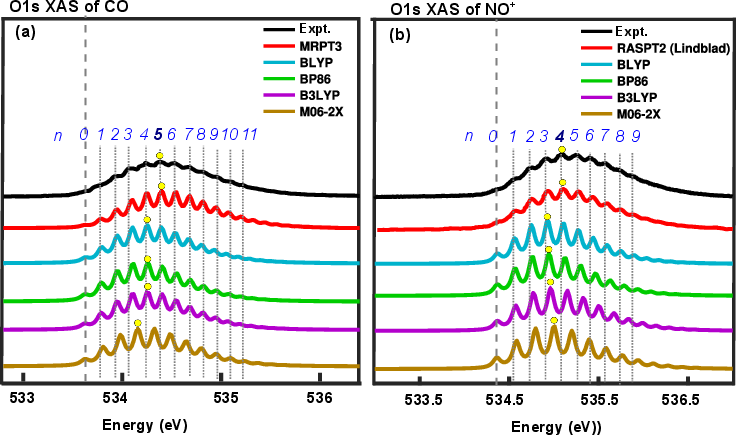}
\caption{Close-up view of Fig. \ref{fig:conop:td:xas} with only the anharmonic results and experiments. The experimental spectra\cite{puttner_vibrationally_1999,lindblad_experimental_2022} are interpreted with the final-state quantum number $n$. Dashed and dotted lines are to guide eyes. In each experiment, for the peak with the largest intensity, the corresponding  $n$  (i.e., $n_\text{max}$)  is bolded  ($n_\text{max}$=5 in panel a and 4 in panel b). Yellow circles label $n_\text{max}$ predicted by each theoretical method. 
}
\label{fig:conop:td:xas:max}
\end{figure*}

Firstly, we compare the transition energies. For example, when comparing neutral N$_2$ [at 400.87 eV; see Fig. \ref{fig:xas:n2:n2+}(a)] to N$_2^+$ [at 394.29 eV; see Fig. \ref{fig:xas:n2:n2+}(b)], we observed a red shift of -6.58 eV in the experimental 0-0 peak. Similarly, ionizing CO [Fig. \ref{fig:xas:co:co+} (a-b)] to CO$^+$ [Fig. \ref{fig:xas:co:co+} (c-d)] resulted in a red shift of -5.25 eV at the C1s edge and -5.20 eV at the O1s edge. On the other hand, the transition from neutral NO [Fig. \ref{fig:xas:no:no+} (a-b)] to its ionized counterpart NO$^+$ [Fig. \ref{fig:xas:no:no+} (c-d)] exhibited a blue shift of +3.14 eV at the N1s edge and +2.88 eV at the O1s edge. These shifts in XAS transition energy induced by valence ionization appear to be dependent on the parity of the system.

Secondly, we compare the profiles. Valence shell ionization can significantly change the profiles in some cases (N$_2$ and N$_2^+$, Fig. \ref{fig:xas:n2:n2+}; CO and CO$^+$, Fig. \ref{fig:xas:co:co+}) or weakly (NO and NO$^+$, Fig. \ref{fig:xas:no:no+}) change the profiles. The specific profile types are discussed in the next subsection.

Lastly, we noticed that within a molecule-cation pair, achieving the SCF convergence is potentially easier for one system (CO$^+$/NO$^+$/N$_2$) than the other (CO/NO/N$_2^+$), suggesting the strong influence of the change in one electron.  However, this technical behavior is system-dependent and does not have a fixed correlation with net charge or parity.

\subsection{Vibronic profile types}
Two extreme vibronic features were observed: (1) Type I, characterized by dominant 0-0 and 0-1 transitions, and (2) Type II, characterized by numerous weak 0-$n$ features (0-0 transition does not dominate anymore). Type I corresponds to smaller changes in bond lengths and vibrational frequencies, while Type II corresponds to larger changes.

The O1s XAS spectra of NO, NO$^+$, and CO exhibit typical Type II features, corresponding to the largest structural and frequency changes among all examples (Fig. \ref{fig:table1}). On the other hand, the N1s XAS of N$_2^+$, C1s XAS of CO, C/O 1s XAS of CO$^+$, C/O 1s XPS of CO, and N/O 1s XPS of NO can be classified as Type I features, referring to smaller structural and frequency changes. Other spectra fall between these two types.

\subsection{Structure-spectroscopy relation associated with different C/N/O core centers}

From the above analysis, it appears that introducing an O1s core hole will lead to larger structural relaxation than the C/N 1s core holes. This can be attributed to the stronger Coulomb attraction between an oxygen nucleus and its 1s electrons compared to a carbon/nitrogen nucleus. Exciting an O1s electron is more likely to cause significant electronic and nuclear rearrangements.

In the case of CO and CO$^+$,  we observed that C1s excitation resulted in C-O distance changes of 0.025 and 0.037 {\AA}, respectively; while O1s excitation led to changes of 0.163 and 0.047  {\AA}, respectively (Table \ref{tab:bond}). Similar trends were observed in NO and NO$^+$, where O1s excitation resulted in larger structural change compared to N1s excitation within the same system.  Among all systems, bond length changes due to N1s core excitations are typically larger than those observed for C1s, but smaller than those for O1s, although there were exceptions as well (C1s excitation of CO$^+$ and N1s of N$_2^+$, which induced structural changes of similar magnitudes). It would be intriguing to conduct further investigations within the same system, such as CN, CN$^-$, or CN$^+$, in order to compare the structural changes induced by C and N 1s excitations within a single system. 

Concerning XPS, this conclusion also holds true for NO, while the opposite is observed for CO.

\subsection{Multiple electronic excited states}
In this work, we considered only the lowest excited state for each system, which is suitable for XPS where only one core-ionized state is needed. Successful calculations have been achieved in more complex polyatomic systems with one or multiple core centers by combining the DFT-FCH method with Franck-Condon simulations that include the Duschinsky rotation effect \cite{hua_theoretical_2020, wei_vibronic_2022, wei_vibronic_2023, wei_vibronic_2023_iii, cheng_vibrationally-resolved_2022}.

However, for XAS, although the lowest core-excited state is often well separated from other core-excited states \cite{hua_jpcl_2019}, some systems may have multiple states that are close in energy. In our study, all systems contained a well-separated lowest core-excited state, except for NO. The experimental O1s XAS spectrum of NO \cite{puttner_vibrationally_1999} exhibits a broad peak composed of three states: $^2\Sigma^-$ (centered at approximately 532.3 eV), $^2\Delta$ (533.2 eV), and $^2\Sigma^+$ (534.2 eV). Similarly, the N1s XAS spectrum of NO \cite{remmers_high-resolution_1993, wang_filtering_2001} shows a broad peak consisting of $^2\Delta$ (centered at approximately 399.4 eV), $^2\Sigma^-$ (399.7 eV), and $^2\Sigma^+$ (400.0 eV). The DFT-XCH method, with one 1s electron removed or excited to the LUMO, is capable of predicting these states. However, achieving higher-lying excited states (i.e., excitations to LUMO+1, LUMO+2, $\cdots$) can be challenging due to the difficulty of the SCF procedure and the potential for variational collapse.

To approximate the contributions of multiple states without directly accessing them, the linear coupling model (LCM)\cite{macak_simulations_2000} can be used. Within the framework of DFT, previous FCH and HCH calculations on PAH molecules phenanthrene (C$_{14}$H$_{10}$) and coronene (C$_{24}$H$_{12}$) \cite{fronzoni_vibrationally_2014} showed good agreement in C1s XAS with experiments by assuming the harmonic approximation.

In cases where both anharmonicity and multiple states need to be considered, the DFT method can be replaced with multiconfigurational methods such as state-averaged RASSCF or multi-state RASPT2 \cite{zhang_nonlinear_2016}.

\section{Conclusions and Outlook}

To summarize, we have conducted a comprehensive theoretical investigation of vibrationally-resolved X-ray spectra for 6 common diatomic systems (N$_2$, N$_2^+$; NO, NO$^+$; CO, CO$^+$) using both the harmonic oscillator (HO) approximation and anharmonic (AH) methods, covering 14 XAS/XPS spectra of the lowest core excited/ionized state at the C/N/O K-edges. Four different functionals (BLYP, BP86,  B3LYP, and M06-2X) were compared with the multiconfigurational methods. Our calculations highlight the significance of anharmonicity in several systems and demonstrate the accuracy and efficiency of DFT in predicting PECs and vibronic fine structures.

This work represents the first comprehensive investigation of vibrationally-resolved X-ray spectral study for multiple diatomic systems.  By providing rich data on the same footing of theory, the effects of charge states, functional sensitivity, XAS and XPS, and structural changes associated with different core centers, are extensively explored. The anharmonicity and functional sensitivity were found to be more pronounced in diatomic systems compared to polyatomic systems. The centrality of anharmonicity was also previously evidenced in vibrationally-resolved X-ray \cite{moitra_vibrationally_2020} and ultraviolet-visible (UV) absorption spectra,\cite{madsen_vibrationally_2019} as well as vibrational (infrared and Raman) spectra.\cite{brauer_vibrational_2005, panek_anharmonic_2016} Besides, we also found connections to the odd or even number of electrons of the systems and performed statistical analysis, leaving room for further exploration of the underlying physical reasons.

\begin{acknowledgments}
Part of this work was inspired by a previous collaborative study on NO$^+$, and we thank Dr. Rebecka Lindblad for sending us the raw experimental XAS spectra of NO$^+$ and Prof. Hans {\AA}gren and Dr. Rafael Couto for helpful discussions. We also appreciate valuable discussions with Profs. Chen Jing and Guangjun Tian. Financial support from the National Natural Science Foundation of China (Grant No. 12274229) is greatly acknowledged.
\end{acknowledgments}


%



\clearpage
\begin{sidewaystable}
\centering
\begin{threeparttable}
\caption{\vskip 3.0in Table I. Comparison of bond lengths (in \AA) in the optimized ground ($R'$) and 1s excited/ionized ($R$) states of all diatomic systems predicted by DFT with different functionals. 
}\label{tab:bond}
\begin{ruledtabular}
\begin{tabular}{lccccccccccccccc}
Compound &&Expt.&&& BLYP& && BP86& && B3LYP& && M06-2X\\ 
\cmidrule(lr){2-4} \cmidrule(lr){5-7} \cmidrule(lr){8-10} \cmidrule(lr){11-13} \cmidrule(lr){14-16}
&$R'$&$R$&$\Delta R$\tnote{a}&$R'$&$R$&$\Delta R$&$R'$&$R$&$\Delta R$&$R'$&$R$&$\Delta R$&$R'$&$R$&$\Delta R$   \\
\hline
$\text {N$_2$}$     
&1.098\tnote{e, f}&1.164\tnote{f, j}&+0.066&1.103&1.167&+0.064&1.103&1.167&+0.063&1.092&1.153&+0.061&1.087&1.139&+0.053
    \\
$\text {N$_2^+$}$    
&1.116\tnote{f}&1.076&-0.040\tnote{n}&1.116&1.075&-0.041&1.115&1.074&-0.041&1.105&1.063&-0.042&1.099&1.052&-0.047
      \\
$\text {NO(N)}$\tnote{b}  
&1.151\tnote{g, h}&1.214\tnote{g, k}&+0.063&1.163&1.220&+0.057&1.160&1.213&+0.053&1.153&1.215&+0.061&1.154&1.219&+0.065
      \\
$\text {NO(O)}$  
&1.151&1.307\tnote{g}&+0.156&1.163&1.328&+0.165&1.160&1.316&+0.156&1.153&1.319&+0.166&1.154&1.322&+0.168  
      \\
$\text {NO$^+$(N)}$
&1.063\tnote{i}&1.123&+0.060\tnote{o}&1.072&1.129&+0.058&1.071&1.125&+0.055&1.057&1.117&+0.059&1.050&1.109&+0.059
      \\
$\text {NO$^+$(O)}$  
&1.063&1.203&+0.140\tnote{o}&1.072&1.211&+0.139&1.071&1.205&+0.134&1.057&1.199&+0.141&1.050&1.193&+0.142  
      \\
CO(C)              
&1.128\tnote{f, h}&1.153\tnote{f}&+0.025&1.138&1.164&+0.026&1.138&1.161&+0.023&1.126&1.156&+0.029&1.122&1.155&+0.034  
    \\
CO(O)              
&1.128&1.291\tnote{h}&+0.163&1.138&1.300&+0.162&1.138&1.295&+0.157&1.126&1.292&+0.165&1.122&1.293&+0.172  
    \\
$\text{CO$^+$(C)}$ 
&1.115\tnote{f}&1.078&-0.037\tnote{p}&1.125&1.080&-0.045&1.124&1.080&-0.044&1.110&1.068&-0.043&1.102&1.059&-0.044 
    \\
$\text{CO$^+$(O)}$
&1.115&1.162&+0.047\tnote{p}&1.125&1.176&+0.051&1.124&1.175&+0.051&1.110&1.163&+0.052&1.102&1.156&+0.054
    \\
NO(N)[XPS]\tnote{c}        
&1.151&1.129\tnote{l}&-0.022&1.163&1.129&-0.034&1.160&1.125&-0.035&1.153&1.117&-0.036&1.154&1.109&-0.045
    \\
NO(O)[XPS]         
&1.151&1.193\tnote{m}&+0.042&1.163&1.211&+0.048&1.160&1.205&+0.045&1.153&1.199&+0.046&1.154&1.193&+0.039
\\
CO(C)[XPS]       
&1.128&1.063&-0.065\tnote{q}&1.138&1.080&-0.058&1.138&1.080&-0.059&1.126&1.068&-0.059&1.122&1.059&-0.063
    \\
CO(O)[XPS]         
&1.128&1.165&+0.037\tnote{q}&1.138&1.176&+0.038&1.138&1.175&+0.037&1.126&1.163&+0.036&1.122&1.156&+0.035
    \\
MAD\tnote{d} 
& -- & -- & -- &0.009&0.009&0.004&0.008&0.006&0.005&0.004&0.006&0.004&0.008&0.011&0.007
    \\
MAX\tnote{d} 
& -- & -- & -- &0.012&0.021&0.012&0.010&0.017&0.013&0.011&0.013&0.014&0.017&0.025&0.023
\end{tabular}
\end{ruledtabular}
\begin{tablenotes}
\item[a] ${\Delta}R\equiv R-R'$, changes of bond length in \AA. 
\item[b] Core excitation/ionization center specified in parenthesis.
\item[c] [XPS] denotes the comparison of initial and final states involved in XPS. The rest are for XAS.
\item[d] MAD, mean absolute deviation; MAX, maximum absolute deviation.
\item[e] Ehara et al.\cite{ehara_symmetry-dependent_2006}
\item[f] Neeb et al.\cite{neeb_coherent_1994}
\item[g] Remmers et al.\cite{remmers_high-resolution_1993}
\item[h] Püttner et al.\cite{puttner_vibrationally_1999}
\item[i] Albritton et al.\cite{albritton_potential_1979}
\item[j] Chen et al.\cite{chen_k_1989}
\item[k] Wang et al.\cite{wang_filtering_2001}
\item[l] Hoshino et al.\cite{hoshino_vibrationally_2008}
\item[m] Thulstrup et al.\cite{thulstrup_configuration_1972}
\item[n] Lindblad et al.\cite{lindblad_x-ray_2020}
\item[o] Lindblad et al.\cite{lindblad_experimental_2022}
\item[p] Couto et al.\cite{couto_carbon_2020}
\item[q] Hergenhahn.\cite{hergenhahn_vibrational_2004}
\end{tablenotes}   
\end{threeparttable}
\end{sidewaystable}

\clearpage
\begin{sidewaystable}[]
\centering
\caption{\vskip 3.0in Table II. Comparison of vibrational frequencies (in cm$^{-1}$) in the optimized ground ($\omega'$) and 1s excited/ionized ($\omega$) states of all diatomic systems predicted by DFT with different functionals. 
}\label{tab:frequency}
\begin{threeparttable}
\begin{ruledtabular}
\begin{tabular}{lccccccccccccccc}
Compound &&Expt.&&& BLYP &&& BP86& && B3LYP& && M06-2X  \\ 
\cmidrule(lr){2-4} \cmidrule(lr){5-7} \cmidrule(lr){8-10} \cmidrule(lr){11-13} \cmidrule(lr){14-16}
&$\omega'$&$\omega$&$\Delta\omega$\tnote{a}&$\omega'$&$\omega$&$\Delta\omega$ &$\omega'$&$\omega$&$\Delta\omega$&$\omega'$&$\omega$&$\Delta\omega$&$\omega'$&$\omega$&$\Delta\omega$   \\
\hline
$\text {N$_2$}$      
&2359.0\tnote{e, f}&1966.2\tnote{f}&-392.8&2334.9&1925.7&-409.2&2345.7&1953.8&-392.0&2449.1&2037.0&-412.0&2522.6&2140.4&-382.2
     \\
$\text {N$_2^+$}$    
&2207.4\tnote{f, g}&2420.8\tnote{g}&+213.4&2234.1&2436.5&+202.3&2253.3&2449.0&+195.7&2330.2&2565.6&+235.4&2407.2&2707.1&+299.9
      \\
$\text {NO(N)}$\tnote{b}  
&1902.7\tnote{h, i}&1613.1\tnote{k}&-289.6&1848.2&1608.7&-239.6&1878.3&1651.5&-226.9&1926.7&1575.2&-351.5&1959.3&1674.4&-284.9
      \\
$\text {NO(O)}$  
&1902.7&1266.3\tnote{k}&-636.4&1848.2&1168.6&-679.6&1878.3&1215.1&-663.3&1926.7&1196.9&-729.8&1959.3&1193.6&-765.7
       \\
$\text {NO$^+$(N)}$
&2376.7\tnote{j}&1954.3\tnote{l}&-422.4&2333.3&1987.8&-345.6&2353.6&2027.3&-326.2&2479.1&2077.5&-401.6&2587.6&2226.7&-360.9
      \\
$\text {NO$^+$(O)}$  
&2376.7&1534.9\tnote{l}&-841.8&2333.3&1512.6&-820.8&2353.6&1553.1&-800.5&2479.1&1584.1&-895.0&2587.6&1647.4&-940.2
       \\
CO(C)              
&2169.8\tnote{f, i}&2083.6\tnote{f}&-86.2&2114.5&2015.8&-98.8&2121.0&2042.6&-78.4&2210.9&2091.3&-119.7&2273.9&2130.4&-143.5 
    \\
CO(O)              
&2169.8&1338.9\tnote{i}&-830.9&2114.5&1264.8&-849.7&2121.0&1292.1&-828.9&2210.9&1302.0&-908.9&2273.9&1304.0&-969.9 
    \\
$\text{CO$^+$(C)}$ 
&2214.2\tnote{f}&2507.6\tnote{m}&+293.4&2172.5&2478.6&+306.1&2192.0\tnote{p}&2490.0&+298.0&2289.8&2612.5&+322.7&2393.3&2748.2&+354.9
    \\
$\text{CO$^+$(O)}$ 
&2214.2&1817.2\tnote{m}&-397.0&2172.5&1779.0&-393.5&2192.0&1787.9&-404.1&2289.8&1875.3&-414.6&2393.3&1939.3&-454.0
    \\
NO(N)[XPS]\tnote{c}         
&1902.7&2064.8\tnote{j}&+162.1&1848.2&1512.6&-335.7&1878.3&2027.3&+149.0&1926.7&2077.5&+150.7&1959.3&2226.7&+267.5
    \\
NO(O)[XPS]         
&1902.7&1609.0\tnote{h}&-293.7&1848.2&1987.8&+139.5&1878.3&1553.1&-325.3&1926.7&1584.1&-342.7&1959.3&1647.4&-311.9
     \\
CO(C)[XPS]       
&2169.8&2451.9\tnote{o}&+282.1&2114.5&2478.6&+364.1&2121.0&2490.0&+368.9&2210.9&2612.5&+401.6&2273.9&2748.2&+474.4
    \\
CO(O)[XPS]         
&2169.8&1822.8\tnote{o}&-347.0&2114.5&1779.0&-335.5&2121.0&1787.9&-333.1&2210.9&1875.3&-335.6&2273.9&1939.3&-334.5
     \\
MAD\tnote{d}
& -- & -- & -- &47.2&47.7 &30.3 &31.6 &37.3 &29.5 &59.2 &68.1 &44.3&127.6&145.5 &73.9\\
MAX\tnote{d}
& -- & -- & -- &55.3&97.7&82.0&48.8 &73.0 &96.2&122.8&160.6&119.5&210.9&296.3&192.3
\end{tabular}
\end{ruledtabular}
\begin{tablenotes}
\item[a] $\Delta\omega \equiv \omega-\omega'$, changes of frequency in cm$^{-1}$.
\item[b] Core excitation/ionization center specified in parenthesis.  
\item[c] [XPS] denotes the comparison of initial and final states involved in XPS. The rest are for XAS.
\item[d] MAD, mean absolute deviation; MAX, maximum absolute deviation.
\item[e] Ehara et al.\cite{ehara_symmetry-dependent_2006}
\item[f] Neeb et al.\cite{neeb_coherent_1994}
\item[g] Laher et al.\cite{laher_improved_1991}
\item[h] Hoshino et al.\cite{hoshino_vibrationally_2008}
\item[i] P\"{u}ttner et al.\cite{puttner_vibrationally_1999}
\item[j] Albritton et al.\cite{albritton_potential_1979}
\item[k] Remmers et al.\cite{remmers_high-resolution_1993}
\item[l] Lindblad et al.\cite{lindblad_experimental_2022}
\item[m] Couto et al.\cite{couto_carbon_2020}
\item[n] Thulstrup et al.\cite{thulstrup_configuration_1972}
\item[o] Hergenhahn.\cite{hergenhahn_vibrational_2004}
\item[p] By restricted open-shell DFT (RO-DFT; the rest are by unrestricted DFT).
\end{tablenotes}   
\end{threeparttable}
\end{sidewaystable}

\begin{table*}[]
\centering
\caption{Computed adiabatic or vertical  1s ionization potentials ($I^\text{vert}$, $I^\text{ad}$) or 1s excitation energy  ($X^\text{vert}$, $X^\text{ad}$; to the corresponding lowest 1s excited states). DFT with different functionals were used. Comparisons are made with experiments. All energies are in eV.} \label{tab:energies}
\begin{threeparttable}
\begin{ruledtabular}
    \begin{tabular}{lccccccccc}
Compound &&&Adiabatic&&& &Vertical&&
    \\ 
\cmidrule(lr){3-6} \cmidrule(lr){7-10}
  &Expt.&BLYP &BP86& B3LYP& M06-2X&BLYP &BP86& B3LYP& M06-2X  \\
 \hline
$\text {N$_2$}$      
&400.87\tnote{c, d} &400.09&399.90&400.13&400.26&400.31&400.10&400.36&400.45
     \\
$\text {N$_2^+$}$    
&394.29\tnote{e} & 394.39&394.17&394.12&393.78&394.50&394.28&394.25&393.96
      \\
$\text {NO(N)}$\tnote{a}  
&399.39\tnote{f, g} &398.65 &398.48 &399.61 &401.05&398.81&398.63&399.74&401.07
      \\
$\text {NO(O)}$  
&531.48\tnote{h} &531.26 &531.22 &532.05 &533.25&532.11&532.02&532.63&533.84 
       \\
$\text {NO$^+$(N)}$  
&402.53\tnote{i} &401.06 &400.86 &402.08 &403.62&401.28&401.06&402.29&403.79
      \\
$\text {NO$^+$(O)}$  
&534.36\tnote{i} &533.22 &533.10 &534.00 &535.35&534.11&533.96&534.94&536.27
       \\
CO(C)              
&287.29\tnote{d} &286.08 &285.81 &286.98 & 288.27&286.11&285.84&287.01&288.27
    \\
CO(O)              
&533.62\tnote{h} &532.42 &532.29 &533.01 &533.99&533.24&533.07&533.85&534.79
    \\
$\text{CO$^+$(C)}$ 
&282.04\tnote{j} &282.16 &281.84 &282.19 &282.15&282.29&281.97&282.33&282.32
    \\
$\text{CO$^+$(O)}$ 
&528.42\tnote{j} &528.26 &528.12 &528.04 &527.66&528.38&528.24&528.18&527.82
    \\
NO(N)[XPS]\tnote{b}         
&410.19\tnote{k, l} &410.78 &410.68 &411.14 &411.31&410.83&410.75&411.24&411.51
    \\
NO(O)[XPS]         
&543.20\tnote{m} &542.93 &542.93 &543.07 &543.03&543.01&543.00&543.12&543.04
\\
CO(C)[XPS]        
&296.13\tnote{n} &296.02 &295.82 &296.25 &296.26&296.24&296.05&296.50&296.58
    \\
CO(O)[XPS]         
&542.62\tnote{n} &542.12 &542.10 &542.09 &541.76&542.18&542.16&542.16&541.82
    \end{tabular}
    \begin{tablenotes}
\item[a] Core excitation/ionization center is specified in parenthesis, where necessary.
\item[b] [XPS] denotes ionic potentials (in XPS); the rest are excitation energies (in XAS).
\item[c] Chen et al.\cite{chen_k_1989}
\item[d] Hitchcock et al.\cite{hitchcock_k-shell_1980}
\item[e] Lindblad et al.\cite{lindblad_x-ray_2020}
\item[f] Wang et al.\cite{wang_filtering_2001}
\item[g] Remmers et al.\cite{remmers_high-resolution_1993}
\item[h] P\"{u}ttner et al.\cite{puttner_vibrationally_1999}
\item[i] Lindblad et al.\cite{lindblad_experimental_2022}
\item[j] Couto et al.\cite{couto_carbon_2020}
\item[k] Hoshino et al.\cite{hoshino_vibrationally_2008}
\item[l] Bagus et al.\cite{bagus_anomalous_1974}
\item[m] Bakke et al.\cite{bakke_table_1980}
\item[n] Hergenhahn.\cite{hergenhahn_vibrational_2004}
\end{tablenotes}  
\end{ruledtabular}
\end{threeparttable}
\end{table*}

\clearpage
\begin{table*}[]
    \centering
    \caption{Fitted Morse\cite{morse_diatomic_1929} parameters for the potential energy curves of selected systems (CO$^\text{+}$, NO, CO, and NO$^\text{+}$) in the ground and the lowest 1s excited states by using different electronic structure methods. All parameters are in atomic units.
    } \label{tab:morse}

\begin{ruledtabular}
\begin{threeparttable}
  \begin{tabular}{lccccccccc}
 Method  & $T_\text{e}$\tnote{a} & $D_\text{e}$ & $\alpha$ & $R_\text{e}$ & $T_\text{e}$ & $D_\text{e}$ & $\alpha$ & $R_\text{e}$ &\\ \hline
 &\textbf{CO$^+$ ground}    &         &     &       &    \textbf{NO ground}& & && \\
 RASPT2\tnote{b}  & 0.0023 & 0.3528 & 1.3600 & 2.0947 & --& --& --& --&  \\ 
 RASSCF           & 0.0043 & 0.3206 & 1.4663 & 2.0839 & --& --& --& --&  \\ 
  BLYP             &-0.0002 & 0.3524 & 1.3257 & 2.1210 &-0.0001 &0.2789 &1.3421 &2.1916 & \\
  BP86             &-0.0003 & 0.3660 & 1.3131 & 2.1196 
  &0.0001 &0.2909 &1.3354 &2.1855 & \\
  B3LYP             &0.0015 & 0.3489 & 1.4016 & 2.0847 
  &0.0026 &0.2758 &1.4104 &2.1543 & \\
  M06-2X            &0.0027 & 0.3738 & 1.4132 & 2.0661 
  &0.0036 &0.2923 &1.4223 &2.1340 & \\
  &\textbf{CO$^+$ C1s} &             &   &   & \textbf{NO N1s} & & & & \\
 RASPT2           & 10.3539 & 0.3133 & 1.5421 & 2.0330 & --& --& --& --&  \\
 RASSCF           & 10.4314 & 0.3643 & 1.6221 & 1.9895 & --& --& --& --&  \\ 
  BLYP             &10.3655 & 0.3592 & 1.4681 & 2.0387 &14.6458 &0.1544 &1.4667 &2.3164 & \\
  BP86             &10.3541 & 0.3578 & 1.4820 & 2.0353 &14.6392 &0.1616 &1.4673 &2.3037 & \\
  B3LYP            &10.3679 & 0.3590 & 1.5336 & 2.0109 &14.6819 &0.1527 &1.5252 &2.2819 & \\
  M06-2X           &10.3665 & 0.4558 & 1.4540 & 1.9926 &14.7355 &0.2121 &1.4642 &2.2337 & \\
 & \textbf{CO$^+$ O1s}        &   &   &       & \textbf{NO O1s} & & & & \\
 RASPT2           & 19.4320 & 0.2778 & 1.3205 & 2.2142 & --& --& --& --& \\ 
 RASSCF           & 19.5124 & 0.2855 & 1.4308 & 2.1339 & --& --& --& --& \\ 
  BLYP             &19.4007 & 0.2596 & 1.3102 & 2.2217 &19.5149 &0.0858 &1.4246 &2.5177 & \\
  BP86             &19.3969 & 0.2394 & 1.3607 & 2.2143 &19.5120 &0.0975 &1.4088 &2.4923 & \\
  B3LYP            &19.3918 & 0.2961 & 1.2893 & 2.1978 &19.5431 &0.0973 &1.4294 &2.4745 & \\
  M06-2X           &19.3765 & 0.3481 & 1.2336 & 2.1899 &19.5880 &0.1147 &1.4267 &2.4291 & \\
& \textbf{CO ground}           &   &   &    &  \textbf{NO$^+$ ground}  & & & & \\
 MRPT3\tnote{c}  & 0.0018 & 0.4239 & 1.2168 & 2.1346 &-- & --&-- &-- & \\ 
  BLYP             &-0.0031 & 0.4149 & 1.2078 & 2.1560 &-0.0021 & 0.4695 & 1.2793 & 2.0388 & \\ 
  BP86             &-0.0032 & 0.4272 & 1.1961 & 2.1565 &-0.0022 & 0.4884 & 1.2657 & 2.0372 & \\
  B3LYP            &-0.0033 & 0.4791 & 1.1739 & 2.1360 &-0.0015 & 0.5644 & 1.2339 & 2.0132 &\\
  M06-2X           &-0.0035 & 0.5436 & 1.1345 & 2.1288 &-0.0014 & 0.6730 & 1.1780 & 2.0004 & \\
& \textbf{CO O1s}        &   &     &  & \textbf{NO$^+$ O1s}  & & & & \\ 
 MRPT3             &19.6227 & 0.1751 & 1.1700 & 2.4475   &-- & --&-- &-- & \\ 
  BLYP             &19.5558 & 0.1407 & 1.3100 & 2.4375  &19.5858 & 0.1591 & 1.4631 & 2.2769 &\\
  BP86             &19.5508 & 0.1460 & 1.3092 & 2.4274  &19.5814 & 0.1693 & 1.4535 & 2.2663 & \\
  B3LYP            &19.5784 & 0.1378 & 1.3528 & 2.4091 &19.6141 & 0.1782 & 1.4643 & 2.2442 & \\
  M06-2X           &19.6100 & 0.1870 & 1.2649 & 2.3850 &19.6623 & 0.2132 & 1.4301 & 2.2176 & \\
    \end{tabular}
    \begin{tablenotes}
\item[a] $U(R)=T_\text{e}+D_\text{e}[1-e^{-\alpha (R-R_\text{e})}]^2$. $R_\text{e}$, equilibrium internuclear distance; $D_\text{e}$, dissociation energy (well depth); $\alpha$ = $\sqrt{k_\text{e}/2D_\text{e}}$, a potential width parameter; $k_\text{e}=[\frac{d^{2}U(R)}{dR^{2}}]_{R=R_\text{e}}$, force constant at $R_\text{e}$; $T_\text{e}$, a constant term.
\item[b] RASPT2 PEC curves taken from Couto et al.\cite{couto_carbon_2020}
\item[c] MRPT3 PECs curves taken from Huang et al.\cite{huang_theoretical_2022}
\end{tablenotes}   
\end{threeparttable}
\end{ruledtabular}
\end{table*}


\end{document}